\RequirePackage{fix-cm}
\documentclass[smallextended]{svjour3}       
\smartqed  

\usepackage{graphicx}
\usepackage{mathtools}
\usepackage{amsmath}
\usepackage{graphicx}
\usepackage{subcaption}
\usepackage{float}        
\usepackage{placeins}     
\usepackage[skip=4pt,font=small]{caption}
\usepackage{float}
\begin{document}
\title{Hawking Radiation and Second-Order Quantum-Corrected Thermodynamics with Black Hole Remnant Formation in a Vaidya-Bonnor Black Hole Surrounded by Quintessence 
}


\author{Djedai Ayang Kamo  	\and Saleh Mahamat      \and Ragil Brand Tsafack Ndongmo    \and  Kamiko Kouemeni Jean Rodrigue       \and Thomas Bouetou Bouetou \and  Timoleon Crepin Kofane
}


\institute{Djedai Ayang Kamo \at
               \
              Department of Physics, Faculty of Science, University of Maroua, P.O. Box. 814, Maroua, Cameroon\\
              \
              \email{sylvaindjedaiayang@gmail.com}           
           \and
           Saleh Mahamat \at
            \
           Research Center in Didactics of Fundamental and Applied Sciences (CERDISFA).\\ 
            Department of Physics, Higher Teachers' Training College, University of Maroua, P.O. Box 55, Maroua, Cameroon\\
              \
        \email{mahsaleh2000@yahoo.fr}           
           \and
           Ragil Brand Tsafack Ndongmo \at
             \
             Department of Physics, Faculty of Science, University of Yaounde I, P.O. Box. 812, Yaound\'e, Cameroon.\\
            Fakult\"at f\"ur Mathematik, Ruhr-Universit\"at Bochum, Universit\"atsstraße 150, 44780, Bochum, Germany, \\
            \
           \email{nragilbrand@gmail.com}
           \and
           Kamiko Kouemeni Jean Rodrigue \at
              \
              Department of Physics, Faculty of Science, University of Maroua, P.O. Box 814 Maroua, Cameroon\\
              \
              \email{lafraise2001@yahoo.fr}           
           \and
           Thomas Bouetou Bouetou \at
              \
              National Advanced School of Engineering, University of Yaound\'e I, P.O. Box. 8390,
Yaounde, Cameroon,\\
              \
              \email{tbouetou@yahoo.fr}           
           \and
            Timoleon Crepin Kofane \at
              \
              Department of Physics and Astronomy, Botswana International University of Science and Technology,
Private Mail Bag 16, Palapye, Botswana\\
              \
              \email{tckofane@gmail.com}           
           \and
}

\date{Received: date / Accepted: date}

\maketitle

\begin{abstract}
In this work, we investigate Hawking radiation and second-order quantum-corrected thermodynamics of a Vaidya–Bonnor black hole surrounded by quintessence. Within the Hamilton–Jacobi tunneling formalism, analytical expressions for the event horizons, tunneling probability, and Hawking temperature are derived by treating the quintessence field as a perturbative contribution. The classical thermodynamic quantities, including entropy, heat capacity, enthalpy, Helmholtz free energy, and Gibbs free energy, are then obtained and analyzed. Quantum thermal fluctuations are incorporated through logarithmic and inverse-entropy corrections to the Bekenstein–Hawking entropy, leading to second-order corrected thermodynamic quantities. Their effects on thermal stability and phase transitions are examined in detail. Furthermore, analytical expressions for the quantum-corrected remnant radius and remnant mass are derived, showing that the combined effects of quintessence and quantum corrections can prevent complete black-hole evaporation and lead to the formation of a stable remnant. Our results show that both quintessence and higher-order quantum corrections enhance thermodynamdymic stability and significantly favor the formation of stable black-hole remnants.
\end{abstract}

\section{Introduction}
Black holes constitute one of the most remarkable predictions of Einstein's general theory of relativity and provide a unique theoretical framework for exploring the interplay between gravitation, quantum mechanics and thermodynamics. For a long time, they were regarded as perfectly absorbing objects from which nothing could escape. This classical picture changed dramatically following the pioneering works of Bekenstein and Hawking, who demonstrated that black holes obey thermodynamic laws analogous to those of ordinary physical systems. Bekenstein established that the entropy of a black hole is proportional to the area of its event horizon, while Hawking proved that quantum effects near the event horizon lead to the emission of thermal radiation with a characteristic temperature, now known as Hawking radiation [1–5]. These groundbreaking discoveries established the foundations of black hole thermodynamics and initiated extensive investigations into the quantum nature of gravity.
Following these seminal developments, numerous approaches have been proposed to investigate Hawking radiation and its physical implications. Among them, the quantum tunneling method has emerged as one of the most successful semiclassical approaches for deriving the Hawking temperature and emission probability. In particular, the Parikh–Wilczek tunneling formalism, together with its subsequent extensions to charged, rotating and fermionic particles, has provided a consistent and physically intuitive description of black hole evaporation while naturally incorporating self-gravitational effects and energy conservation [6–12]. These achievements have considerably improved our understanding of the microscopic origin of Hawking radiation and stimulated further studies of black hole evaporation in both static and dynamical spacetimes.
Unlike stationary black holes, realistic astrophysical black holes are expected to evolve through continuous processes of matter accretion and radiation emission. Consequently, dynamical black hole solutions play a crucial role in describing realistic gravitational systems. Among these solutions, the Vaidya spacetime and its charged extension, namely the Vaidya–Bonnor black hole, provide an appropriate description of radiating charged black holes with a time-dependent mass function [13–16]. Owing to their non-static nature, these geometries have attracted considerable attention in studies of gravitational collapse, Hawking radiation, horizon dynamics and black hole thermodynamics.
On the other hand, compelling cosmological observations indicate that the Universe is currently undergoing an accelerated expansion driven by an exotic component commonly referred to as dark energy. Quintessence, characterized by a negative pressure and an equation-of-state parameter satisfying $-1<\epsilon<-1/3$, represents one of the most widely investigated dynamical models of dark energy[1–16]. When surrounding a black hole, quintessence modifies the spacetime geometry, alters the horizon structure and significantly affects the thermodynamic properties of the system. Consequently, black holes immersed in a quintessence field have been extensively investigated in recent years, revealing significant modifications of Hawking temperature, entropy, critical behaviour and phase transitions [17–30].
Besides these classical aspects, increasing attention has recently been devoted to quantum corrections to black hole thermodynamics. Near the final stages of black hole evaporation, thermal and quantum fluctuations become unavoidable, causing deviations from the classical Bekenstein–Hawking entropy. These effects generally introduce a logarithmic correction as the leading-order contribution, followed by higher-order terms inversely proportional to the entropy. Such corrections modify not only the entropy but also all associated thermodynamic potentials, including the internal energy, Helmholtz free energy, enthalpy, Gibbs free energy and heat capacity. They also provide a possible mechanism for the formation of black hole remnants and may offer valuable insights into the underlying theory of quantum gravity [31–40]. Recent studies have further investigations quantum corrections in charged black hole thermodynamics, showing that quantum fluctuations can significantly modify the Hawking temperature, entropy, heat capacity, critical behavior and remnant formation [41]. In particular, quantum-corrected Reissner–Nordström and Reissner–Nordström–AdS black holes have been shown to exhibit modified critical and remnant states [42]. More recently, the combined effects of quantum corrections and quintessence on the thermodynamic criticality of charged AdS black holes have also been investigated [43]. For dynamical black holes, recent work has emphasized the role of Vaidya radiation in the context of Hawking emission [44].
Futhermore, several investigations have explored quantum-corrected thermodynamics for dynamical and non-static black hole geometries. These studies demonstrated that thermal fluctuations may significantly influence the thermodynamic stability and phase structure of black holes, particularly in the small-horizon regime. Furthermore, recent analyses have extended these corrections to Vaidya-type spacetimes and other black hole configurations, highlighting the important role of higher-order entropy corrections in describing realistic evaporation processes [45–55]. In particular, a recent investigation of a nonlinear magnetically charged AdS black hole surrounded by quintessence showed that second-order entropy corrections considerably modify the entropy, enthalpy, Helmholtz free energy, Gibbs free energy and heat capacity, thereby providing a more complete description of black hole thermodynamics beyond the semiclassical approximation [55].
Despite these remarkable advances, an important gap still exists in the current literature. Most previous investigations have focused either on Hawking radiation from dynamical black holes or on the quantum-corrected thermodynamics of static black hole geometries. Similarly, studies devoted to black holes surrounded by quintessence have mainly concentrated on their classical thermodynamic properties, critical behaviour and phase transitions. Although second-order entropy corrections have recently been investigated for nonlinear magnetically charged AdS black holes immersed in a quintessence background [55], a comprehensive analysis combining Hawking radiation and second-order quantum-corrected thermodynamics for a dynamical Vaidya–Bonnor black hole surrounded by quintessence is still lacking. In particular, the combined influence of the quintessence parameters on the Hawking radiation process, the corrected thermodynamic quantities and the thermal stability of such a dynamical spacetime has not yet been systematically explored. 
Motivated by this gap, the present work investigates the Hawking radiation and the second-order quantum-corrected thermodynamics of a Vaidya-Bonnor black hole surrounded by quintessence. Hawking radiation is derived within the semiclassical quantum tunneling framework, allowing the corresponding Hawking temperature and tunneling probability to be obtained. Building upon the corrected entropy formalism, we derive analytical expressions for the second-order corrected entropy, Hawking temperature, heat capacity, Helmholtz free energy, internal energy, enthalpy and Gibbs free energy. The effects of the quintessence normalization parameter $c$, the equation-of-state parameter $\epsilon$, the electric charge $Q$, and the quantum correction parameters on the thermodynamic behaviour and stability of the black hole are analysed in detail through analytical and graphical investigations.
An additional objective of this work is to examine the possible existence of a black hole remnant during the final stage of evaporation. Since quantum corrections become dominant when the event horizon approaches the Planck scale, they may prevent the complete evaporation of the black hole and lead to the formation of a stable remnant with finite mass and radius. Such remnants have attracted considerable attention because they may provide a possible resolution of the information loss paradox and offer valuable insights into quantum gravity. The possibility of black hole remnant formation has been extensively inestigated in several approaches to quantum gravity, including the generalized uncertainty principle, noncommutative geometry, gravity's rainbow and information-loss scenarios. These studies suggest that quantum gravitational effects may halt the evaporation process, leaving behing a stable remnant with finite mass and radius[35–40, 56-63]. In this work, the remnant radius is determined from the corrected thermodynamic framework, and the influence of the quintessence field on its formation and physical characteristics is investigated.
Unlike previous investigations, the present study combines, within a unified framework, Hawking radiation, second-order quantum-corrected thermodynamics and black hole remnant formation for a dynamical Vaidya–Bonnor black hole surrounded by quintessence. 
The paper is organized as follows. Section 2 presents the geometry of the Vaidya–Bonnor black hole surrounded by quintessence and discusses its horizon structure. Section 3 is devoted to the derivation of Hawking radiation using the quantum tunneling approach. In Section 4, the second-order quantum-corrected thermodynamic quantities are derived analytically, and the existence of the black hole remnant is investigated. Section 5 discusses the physical effects of the model parameters on the corrected thermodynamic quantities and the stability of the black hole. Finally, the main conclusions and future perspectives are summarized in the last section.

\section{Horizon Structure and Hawking Radiation via the Hamilton-Jacobi Method}

We consider a spherically symmetric Vaidya-Bonnor black hole spacetime surrounded by quintessence. In the quasi-static approximation, the slowly varying mass and charge functions are treated as approximately constant during the tunneling[40-55]

\begin{equation}
ds^{2}=-\Psi(r,v)dv^{2}+2dvdr+r^{2}\left(d\theta^{2}+\sin^{2}\theta d\phi^{2}\right),\label{eq1}
\end{equation}

where the metric function takes the form

\begin{equation}
\Psi(r,v)=1-\frac{2M(v)}{r}+\frac{Q^{2}(v)}{r^{2}}-\frac{c}{r^{3\varepsilon+1}}.\label{eq2}
\end{equation}

Here $M(v)$ and $Q(v)$ are the mass and charge functions depending on the advanced null coordinate $v$. In the quasi-static approximation, where the variations of $M(v)$ and $Q(v)$ are sufficiently slow compared with the tunneling process, one may set $M(v)\rightarrow M,\qquad Q(v)\rightarrow Q$, $c$ the normalization factor associated with quintessence, and $\varepsilon$ the quintessence equation-of-state parameter satisfying $-1\leq \varepsilon \leq -\frac{1}{3}$.

For $c=0$, Eq.~(\ref{eq2}) reduces to the Reissner-Nordström black hole, whereas for $Q=0$, it reduces to the Schwarzschild black hole surrounded by quintessence.

The event horizons are determined from the condition

\begin{equation}
\Psi(v,r)=0.
\label{eq7}
\end{equation}

Substituting Eq.~(\ref{eq2}) into Eq.~(\ref{eq7}), we obtain

\begin{equation}
1-\frac{2M(v)}{r}+\frac{Q(v)^{2}}{r^{2}}-\frac{c}{r^{3\varepsilon+1}}=0.
\label{eq8}
\end{equation}

Multiplying Eq.~(\ref{eq8}) by $r^{3\varepsilon+1}$, one gets

\begin{equation}
r^{3\varepsilon+1}-2M(v)r^{3\varepsilon}+Q(v)^{2}r^{3\varepsilon -1}-c=0.
\label{eq9}
\end{equation}

Factoring out $r^{3\varepsilon-1}$, Eq.~(\ref{eq9}) becomes

\begin{equation}
r^{3\varepsilon-1}\left(r^{2}-2M(v)r+Q(v)^{2}\right)-c=0.
\label{eq10}
\end{equation}

In the absence of quintessence $(c=0)$, the horizon equation reduces to

\begin{equation}
r^{2}-2M(v)r+Q(v)^{2}=0.
\label{eq11}
\end{equation}

The corresponding roots are

\begin{equation}
r_{\pm}=M\pm \sqrt{M^{2}-Q^{2}}.
\label{eq12}
\end{equation}

Assuming that the quintessence parameter is small $(c\ll 1)$, the corrected horizons may be obtained perturbatively.

For the outer horizon, we write

\begin{equation}
r_c=r_+ + \delta_+.
\label{eq15}
\end{equation}

Substituting into Eq.~(\ref{eq10}) and expanding around $r=r_+$, we obtain

\begin{equation}
r_+^{3\varepsilon-1}(r_+-r_-)\delta_+-c=0.
\label{eq16}
\end{equation}

Using

\begin{equation}
r_+-r_-=2\sqrt{M^{2}-Q^{2}},
\label{eq17}
\end{equation}

the correction becomes

\begin{equation}
\delta_+ =\frac{c}{2\sqrt{M^{2}-Q^{2}}(M+\sqrt{M^{2}-Q^{2}})^{3\varepsilon-1}}.
\label{eq18}
\end{equation}

Therefore, the outer horizon is

\begin{equation}
r_c=M+\sqrt{M^{2}-Q^{2}}+\frac{c(M+\sqrt{M^{2}-Q^{2}})^{1-3\varepsilon}}{2\sqrt{M^{2}-Q^{2}}}.
\label{eq19}
\end{equation}

Similarly, for the inner horizon

\begin{equation}
r_b=r_-+\delta_-,
\label{eq20}
\end{equation}

we obtain

\begin{equation}
r_b=M-\sqrt{M^{2}-Q^{2}}-\frac{c(M-\sqrt{M^{2}-Q^{2}})^{1-3\varepsilon}}{2\sqrt{M^{2}-Q^{2}}}.
\label{eq21}
\end{equation}

The distance between the horizons is therefore

\begin{equation}
r_c-r_b=2\sqrt{M^{2}-Q^{2}}+\frac{c\left[(M+\sqrt{M^{2}-Q^{2}})^{1-3\varepsilon}+(M-\sqrt{M^{2}-Q^{2}})^{1-3\varepsilon}\right]}{2\sqrt{M^{2}-Q^{2}}}
\label{eq22}
\end{equation}

To investigate the Hawking radiation process, we employ the Hamilton-Jacobi tunneling method. The motion of a scalar particle of mass $m$ in the black-hole background is governed by the Klein-Gordon equation

\begin{equation}
g^{\mu\nu}\partial_{\mu}\Phi\partial_{\nu}\Phi+m^{2}\Phi=0.
\label{eq23}
\end{equation}

Using the WKB approximation,

\begin{equation}
\Phi=\exp\left(\frac{i}{\hbar}I\right),
\label{eq24}
\end{equation}

and taking the semiclassical limit $\hbar\rightarrow0$, Eq.~(\ref{eq23}) reduces to the relativistic Hamilton-Jacobi equation

\begin{equation}
g^{\mu\nu}\partial_{\mu}I\partial_{\nu}I+m^{2}=0.
\label{eq25}
\end{equation}

For the metric (\ref{eq1}), Eq.~(\ref{eq25}) becomes

\begin{equation}
-\frac{1}{\Psi(r)}\left(\frac{\partial I}{\partial t}\right)^2
+\Psi(r)\left(\frac{\partial I}{\partial r}\right)^2
+\frac{1}{r^{2}}\left(\frac{\partial I}{\partial \theta}
\right)^2+\frac{1}{r^{2}\sin^{2}\theta}\left(\frac{\partial I}{\partial \phi}\right)^2+m^{2}=0.
\label{eq26}
\end{equation}

The action is separated as

\begin{equation}
I=-Et+W(r)+J(\theta,\phi),
\label{eq27}
\end{equation}

where $E$ denotes the energy of the emitted particle and $J(\theta,\phi)$ is the angular contribution.

Substituting Eq.~(\ref{eq27}) into Eq.~(\ref{eq26}) yields

\begin{equation}
-\frac{E^{2}}{\Psi(r)}+\Psi(r)\left(\frac{dW}{dr}\right)^2
+\frac{\Lambda}{r^{2}}+m^{2}=0.
\label{eq28}
\end{equation}

where $\Lambda$ is the separation constant.

Solving Eq.~(\ref{eq28}) for the radial function gives

\begin{equation}
\frac{dW}{dr}
=\pm\frac{\sqrt{E^{2}-\Psi(r)\left(m^{2}+\frac{\Lambda}{r^{2}}\right)}}{\Psi(r)}.
\label{eq29}
\end{equation}

Near the event horizon, where

\begin{equation}
\Psi(r_c)=0,
\label{eq30}
\end{equation}

the dominant contribution becomes

\begin{equation}
\frac{dW}{dr}=\pm \frac{E}{\Psi(r)}.
\label{eq31}
\end{equation}

Integrating Eq.~(\ref{eq31}), one obtains

\begin{equation}
W(r)=\pm E \int\frac{dr}{\Psi(r)}.
\label{eq32}
\end{equation}

Expanding the metric function around the event horizon $r=r_c$,

\begin{equation}
\Psi(r)=\Psi'(r_c)(r-r_c)+\mathcal{O}\left[(r-r_c)^2\right],
\label{eq33}
\end{equation}

Eq.~(\ref{eq32}) becomes

\begin{equation}
W(r)=\pm E\int\frac{dr}{\Psi'(r_c)(r-r_c)}.
\label{eq34}
\end{equation}

The integral possesses a simple pole at $r=r_c$. Evaluating it through contour integration yields

\begin{equation}
W(r)=\pm\frac{2i\pi E}{\Psi'(r_c)}.
\label{eq35}
\end{equation}
One solution corresponds to scalar particles moving away from the black hole and the other solution corresponds to particles toward the black hole.

Hence,

\begin{equation}
\mathrm{Im}(W)=\frac{2\pi E}{\Psi'(r_c)}.
\label{eq36}
\end{equation}
Imaginary parts of the action can only come about due to the pole at the horizon.

The tunneling probability is given by [41-54]

\begin{equation}
\Gamma=\exp\left[-2\mathrm{Im}(I)\right]=\exp\left[-2\mathrm{Im}(W)\right].
\label{eq37}
\end{equation}

Taking into account both incoming and outgoing trajectories, one finds

\begin{equation}
\Gamma=\exp\left(-\frac{4\pi E}{\Psi'(r_c)}\right).
\label{eq38}
\end{equation}

Equation~(\ref{eq38}) represents the emission probability of particles crossing the event horizon through quantum tunneling.

Using the metric function, we have

\begin{equation}
\Psi'(r_c)=\frac{1}{r_c}-\frac{Q^2}{{r_c}^3}+\frac{3c\varepsilon}{{r_c}^{3\varepsilon+2}}
\label{eq39}
\end{equation}

Substituting Eq.~(\ref{eq39}) into Eq.~(\ref{eq38}) yields

\begin{equation}
\Gamma=\exp\left[-\frac{4\pi E}{\frac{1}{r_c}-\frac{Q^2}{{r_c}^3}+\frac{3c\varepsilon}{{r_c}^{3\varepsilon+2}}}\right].
\label{eq40}
\end{equation}

Simplifying this  expression, the tunneling probability in terms of the physical parameters $M, Q, c$, and $\varepsilon$ can be written explicitly as

\begin{equation}
\Gamma=\exp\left(-\frac{4\pi E r_c^3}{ r_c^2-Q^2+3c {r_c}^{1-3\epsilon}}\right).
\label{Gamma_SQE}
\end{equation}
where $E$ denotes the energy of the emitted particle, $Q$ is the electric charge, $c$ characterizes the intensity of the quintessence field, $\epsilon$ is its equation-of-state parameter, and $r_c$ represents the horizon radius.

The tunneling probability can be compared with the Boltzmann factor [6-12]

\begin{equation}
\Gamma=\exp\left(-\frac{E}{T_H}\right).
\label{eq43}
\end{equation}

Comparing Eqs.~(\ref{eq38}) and (\ref{eq43}), the Hawking temperature is obtained as

\begin{equation}
T_H=\frac{\Psi'(r_c)}{4\pi}.
\label{eq44}
\end{equation}

\begin{equation}
T_H = \frac{1}{4{\pi r_c}}\left(1-\frac{Q^2}{{r_c}^2}+\frac{3c\varepsilon}{{r_c}^{3\varepsilon+1}}\right).
\end{equation}

This expression confirms that the Hawking temperature is entirely determined by the surface gravity at the event horizon.
The temperature depends explicitly on the black-hole mass, electric charge, quintessence normalization parameter, and equation-of-state parameter. Consequently, the presence of quintessence modifies the horizon structure and directly affects the evaporation process through changes in the thermal radiation spectrum.

\begin{figure}[htbp]
	\centering
	\includegraphics[width=0.75\textwidth]{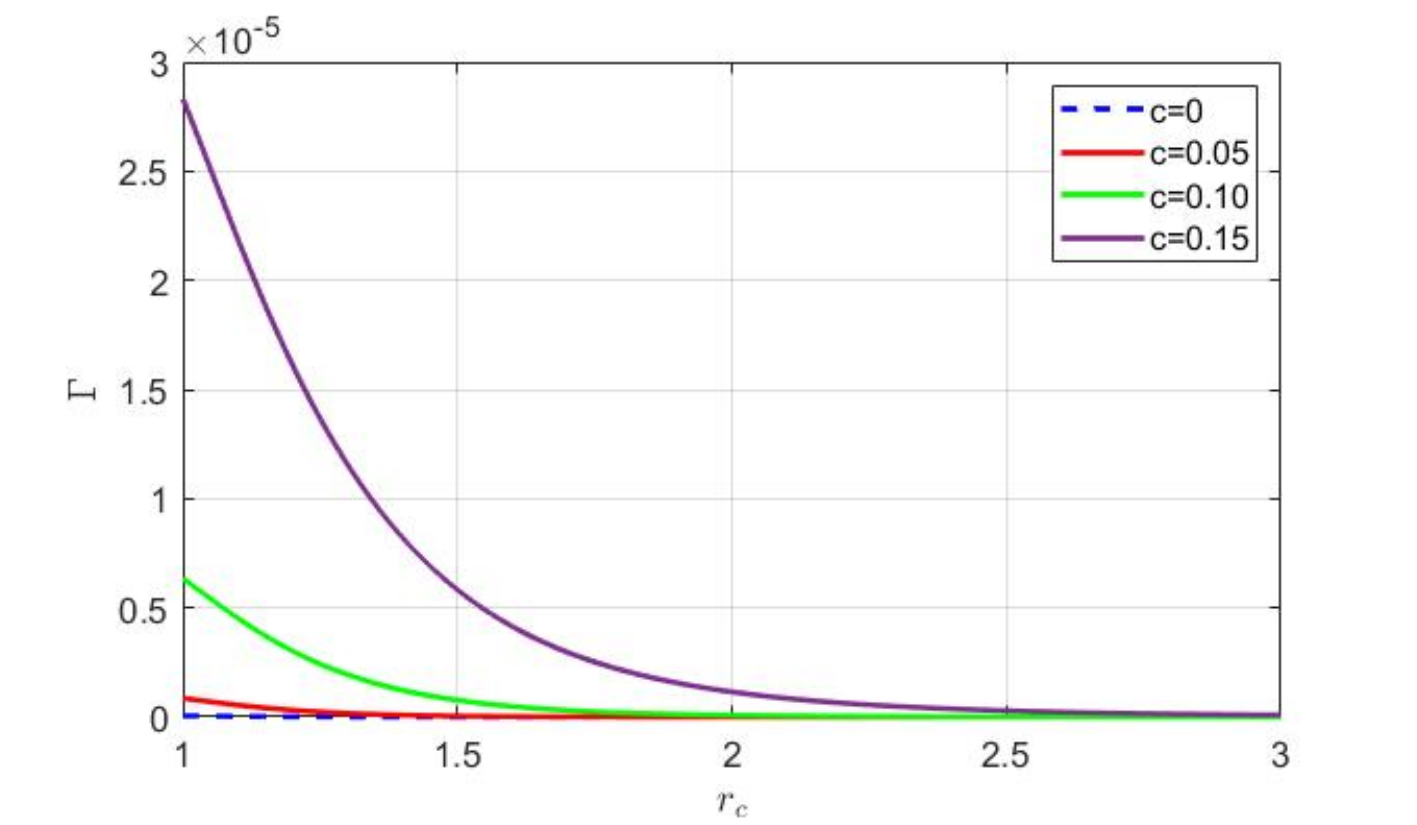}
	\caption{Radiation rate  $\Gamma$ as a function of the horizon radius $r_c$ for different values of the quintessence normalization parameter $c$.}
	\label{fig:T_c}
\end{figure}
Figure~\ref{fig:T_c} shows the effect of the quintessence parameter $c$ on the Hawking radiation rate. For all considered values of $c$, $\Gamma$ decreases monotonically with increasing $r_c$. However, increasing $c$ enhances the radiation rate over the whole range of the horizon radius. This behavior follows directly from Eq.~\eqref{Gamma_SQE}, since the quintessence contribution $3c r_c^{1-3\epsilon}$ increases the denominator of the negative exponent. Consequently, the absolute value of the exponent is reduced and the tunneling probability increases. Thus, as the quintessence parameter $c$ increases, the radiation rate $\Gamma$ also increases.
For $\epsilon=-2/3$, one has $r_c^{1-3\epsilon}=r_c^3$, so that the quintessence contribution becomes particularly important for increasing horizon radius. Physically, the quintessence field modifies the effective gravitational environment surrounding the Vaidya-Bonnor black hole and consequently alters the near-horizon barrier experienced by the emitted particles. The enhancement of $\Gamma$ therefore indicates that, within the considered parameter range, the quintessence background facilitates the tunneling process.

\begin{figure}[htbp]
	\centering	
	\includegraphics[width=0.75\textwidth]{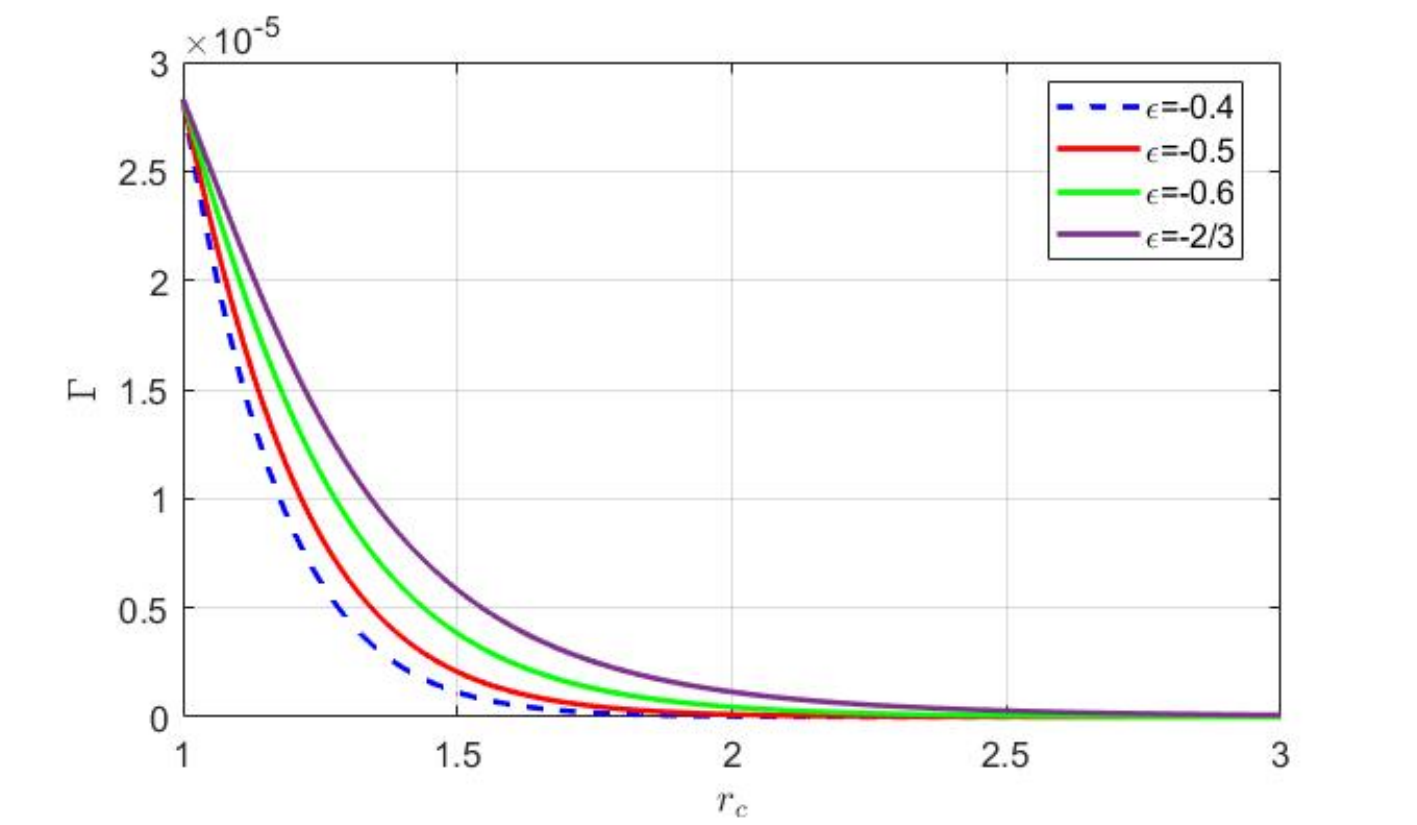}
	\caption{ Radiation rate  $\Gamma$ as a function of the horizon radius $r_c$ for different values of the charge $\epsilon$.}	
	\label{fig:T_epsi}	
\end{figure}
The influence of the equation-of-state parameter $\epsilon$ is presented in Fig.~\ref{fig:T_epsi}. The radiation rate remains a decreasing function of $r_c$ for all values of $\epsilon$, but its magnitude is significantly modified by the quintessence equation of state. Since $\epsilon$ controls the radial dependence of the quintessence contribution through $r_c^{1-3\epsilon}$, changing $\epsilon$ directly modifies the effective near-horizon geometry and, consequently, the tunneling probability. More negative values of $\epsilon$ strengthen the radial contribution of the quintessence term and lead to a larger modification of the radiation rate. Hence, the equation of state of the surrounding dark-energy-like field plays an important role in regulating the Hawking emission. This result emphasizes that the radiation process is sensitive not only to the intensity of the quintessence field but also to its equation of state.
	
\begin{figure}[htbp]
	\centering
	\includegraphics[width=0.75\textwidth]{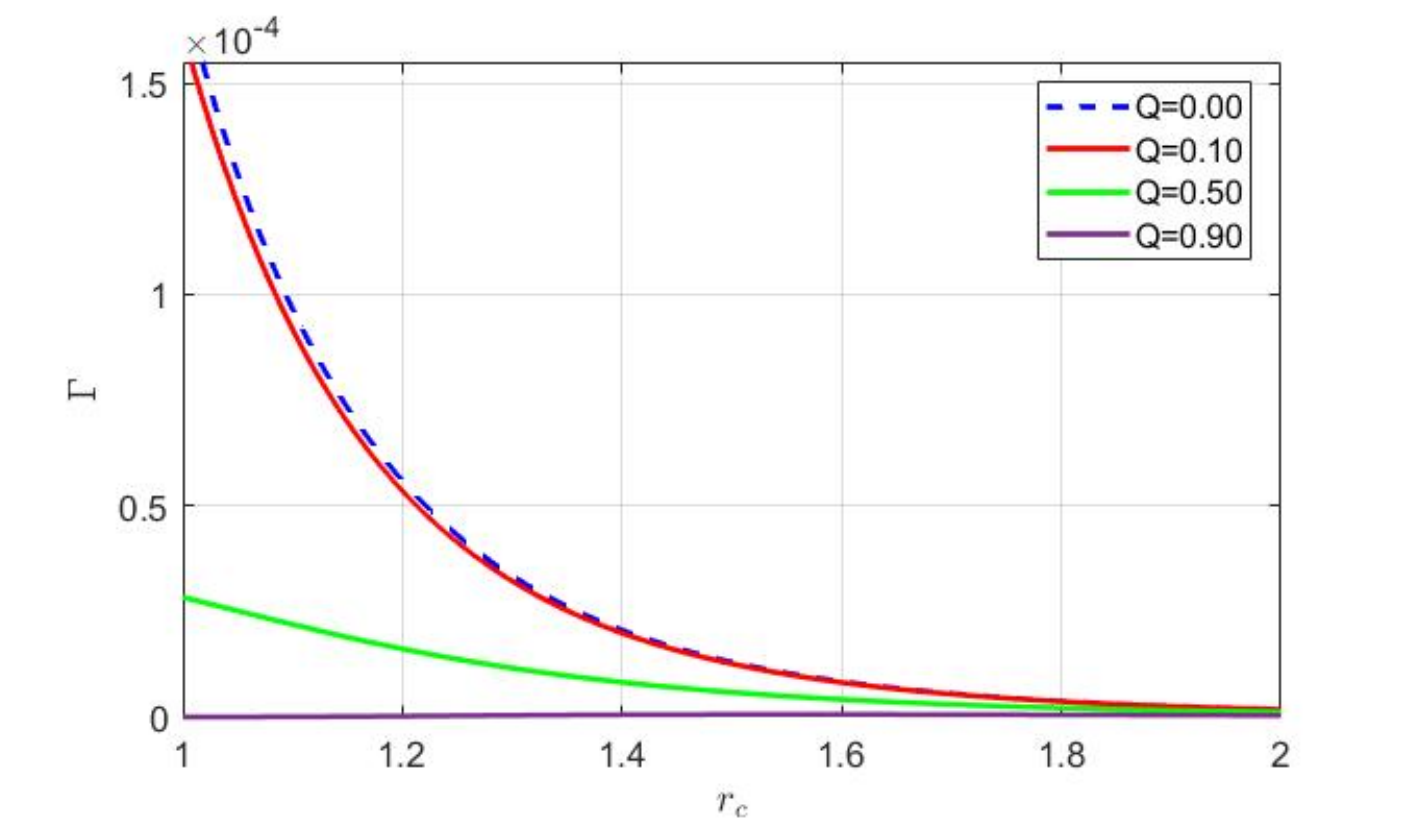}
	\caption{ Radiation rate  $\Gamma$ as a function of the horizon radius $r_c$ for different values of the charge $Q$.}
	\label{fig:T_q}
\end{figure}
Figure~\ref{fig:T_q} illustrates the effect of the electric charge $Q$. A clear suppression of the Hawking radiation rate is observed as the charge increases. In particular, the curves associated with larger values of $Q$ lie below those corresponding to smaller charges, while all curves decrease with increasing $r_c$. This behavior can be understood from the term $-Q^2$ appearing in the denominator of Eq.~\eqref{Gamma_SQE}. Increasing the charge reduces the denominator and consequently increases the magnitude of the negative exponent, leading to a smaller tunneling probability. Therefore, as the charge $Q$ increases, $\Gamma$ decreases.
From a physical point of view, the electromagnetic field modifies the geometry and the effective potential in the vicinity of the horizon, thereby making particle emission less probable. The stronger suppression observed for highly charged configurations reflects the increasing influence of the electromagnetic field on the near-horizon dynamics.

\begin{figure}[htbp]
	\centering
	\includegraphics[width=0.75\textwidth]{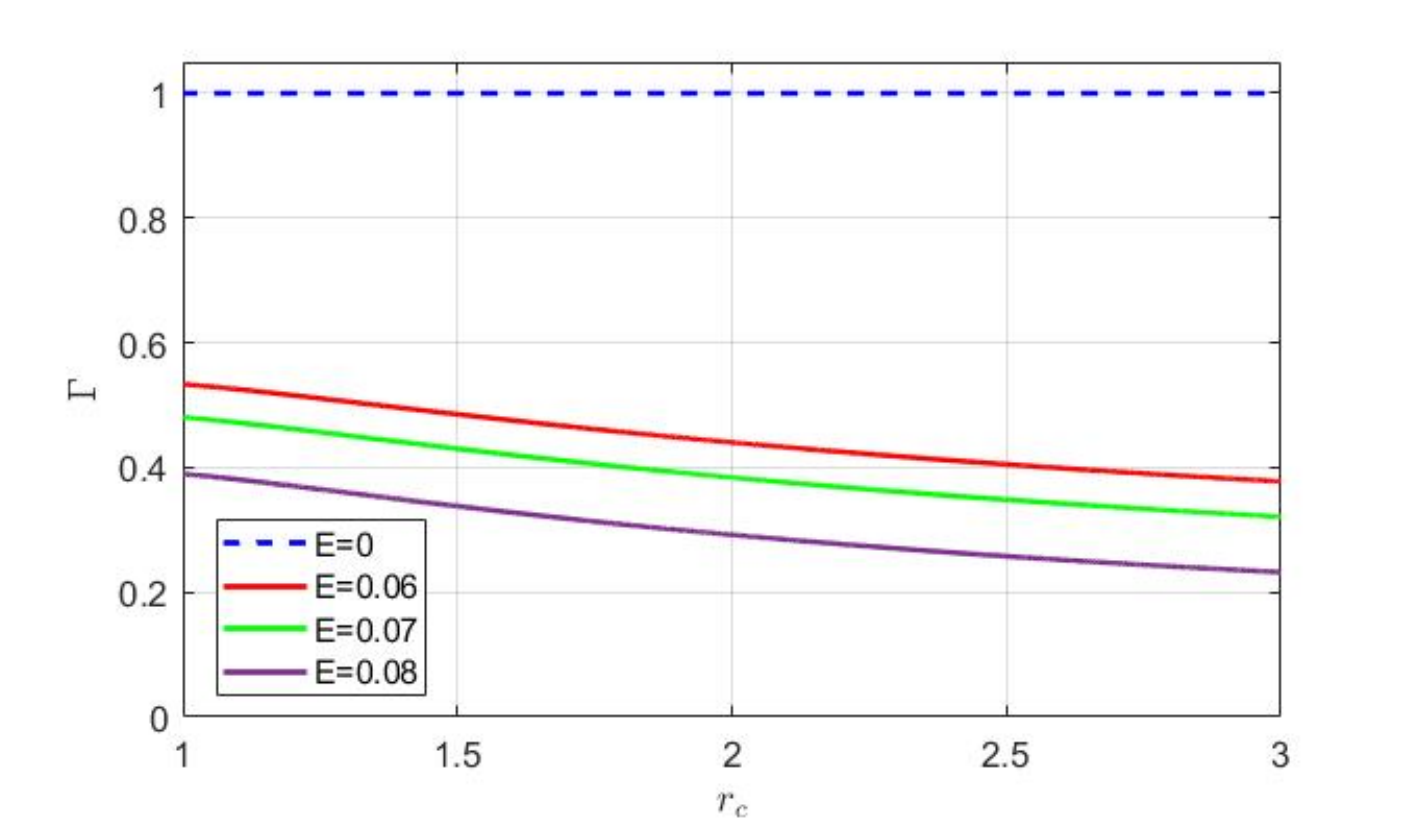}
	\caption{ Radiation rate  $\Gamma$ as a function of the horizon radius $r_c$ for different values of the energy $E$.}
	\label{fig:T_E}
\end{figure}
 Fig.~\ref{fig:T_E} displays the dependence of the radiation rate on the energy $E$ of the emitted particle. The case $E=0$ corresponds to $\Gamma=1$, representing the formal zero-energy limit of Eq.~\eqref{Gamma_SQE}. For non-zero values of $E$, the radiation rate decreases monotonically with increasing $r_c$, and the suppression becomes stronger as the particle energy increases. This behavior is a direct consequence of the exponential dependence of the tunneling probability on $E$. Thus, as the energy $E$ inreases, the radiation rate $\Gamma$ decreases.
Physically, higher-energy particles are associated with a stronger suppression factor in the semiclassical tunneling probability. The increasingly rapid decrease of $\Gamma$ with $E$ therefore reflects the sensitivity of Hawking emission to the energy carried away by the emitted quantum. 

\section{Second-Order Quantum-Corrected Thermodynamics}

\subsection{Second-Order Corrected Entropy}

According to the statistical fluctuation approach, the second-order corrected entropy is expressed as [55]

\begin{equation}
S_{c}=S_{0}-\frac{\alpha}{2}\ln\left(S_{0}T_{H}^{2}\right)+\frac{\beta}{S_{0}},
\end{equation}

where $\alpha$ and $\beta$ are dimensionless correction parameters originating from quantum fluctuations.

For the Vaidya-Bonnor black hole surrounded by quintessence, the Bekenstein-Hawking entropy is

\begin{equation}
S_{0}=\pi r_{c}^{2},
\end{equation}

while the Hawking temperature is

\begin{equation}
T_H = \frac{1}{4{\pi r_c}}\left(1-\frac{Q^2}{{r_c}^2}+\frac{3c\varepsilon}{{r_c}^{3\varepsilon+1}}\right).
\end{equation}

\begin{figure}[htbp]
	\centering
	\includegraphics[width=0.75\textwidth]{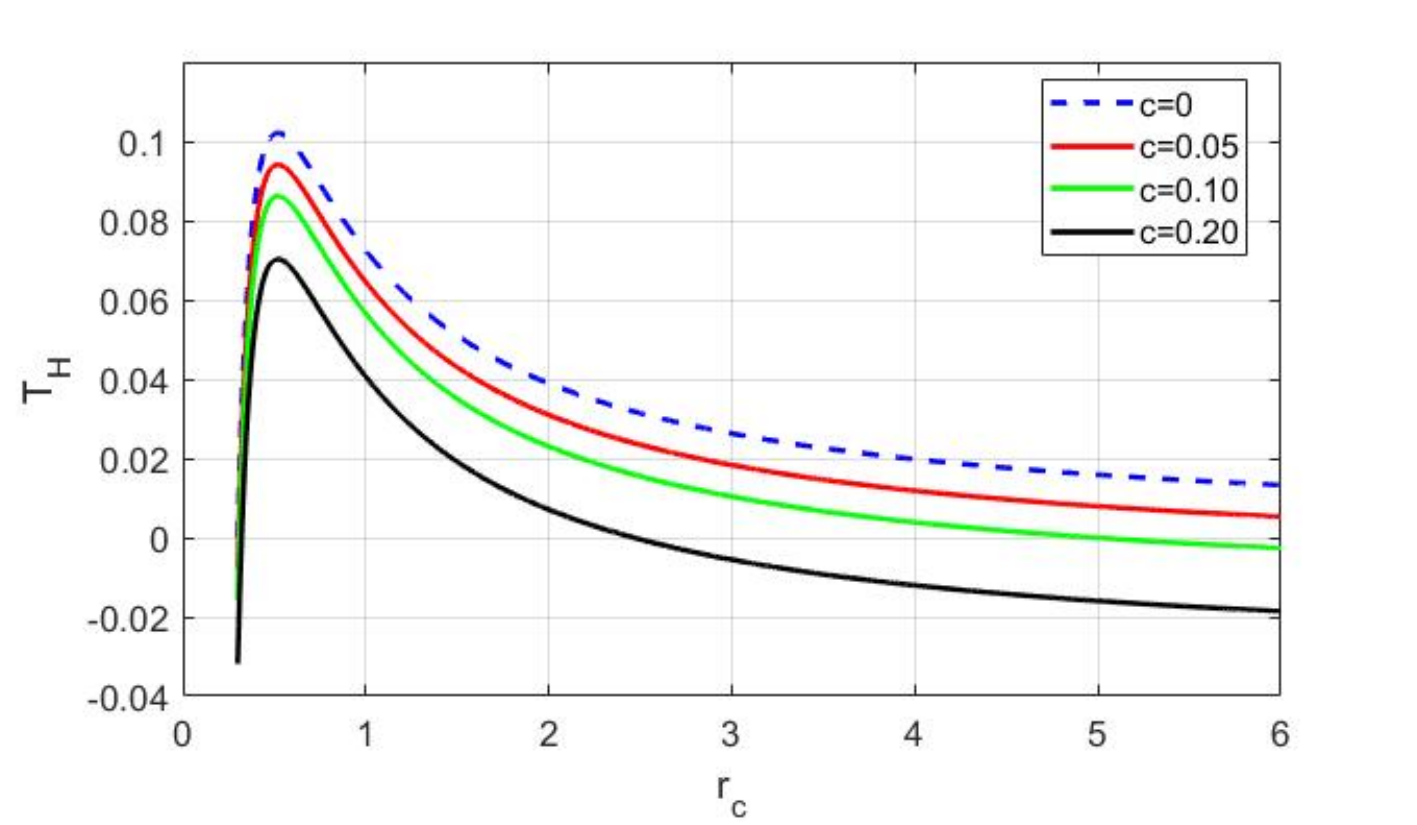}
	\caption{Hawking temperature $T_H$ as a function of the horizon radius $r_c$ for different values of the quintessence normalization parameter $c$.}
	\label{fig:T_H}
\end{figure}

Figure~5 shows that the Hawking temperature is significantly affected by the quintessence parameter $c$. As the value of $c$ increases, the Hawking temperature decreases over the entire range of the event horizon radius. Physically, the quintessence field acts as a dark-energy background characterized by a negative pressure, which modifies the spacetime geometry around the Vaidya-Bonnor black hole. This modification reduces the surface gravity at the event horizon, and consequently lowers the Hawking temperature, since the latter is directly proportional to the surface gravity. The suppression of the Hawking temperature implies that the black hole radiates less efficiently, thereby slowing down the evaporation process. As a result, the presence of a stronger quintessence field tends to enhance the thermodynamic stability of the black hole and promotes the persistence of a finite remnant during the final stage of evaporation. This behavior highlights the significant role played by the surrounding dark-energy environment in regulating the thermal properties and long-term evolution of the black hole.
\bigskip

\begin{figure}[htbp]
	\centering
	\includegraphics[width=0.75\textwidth]{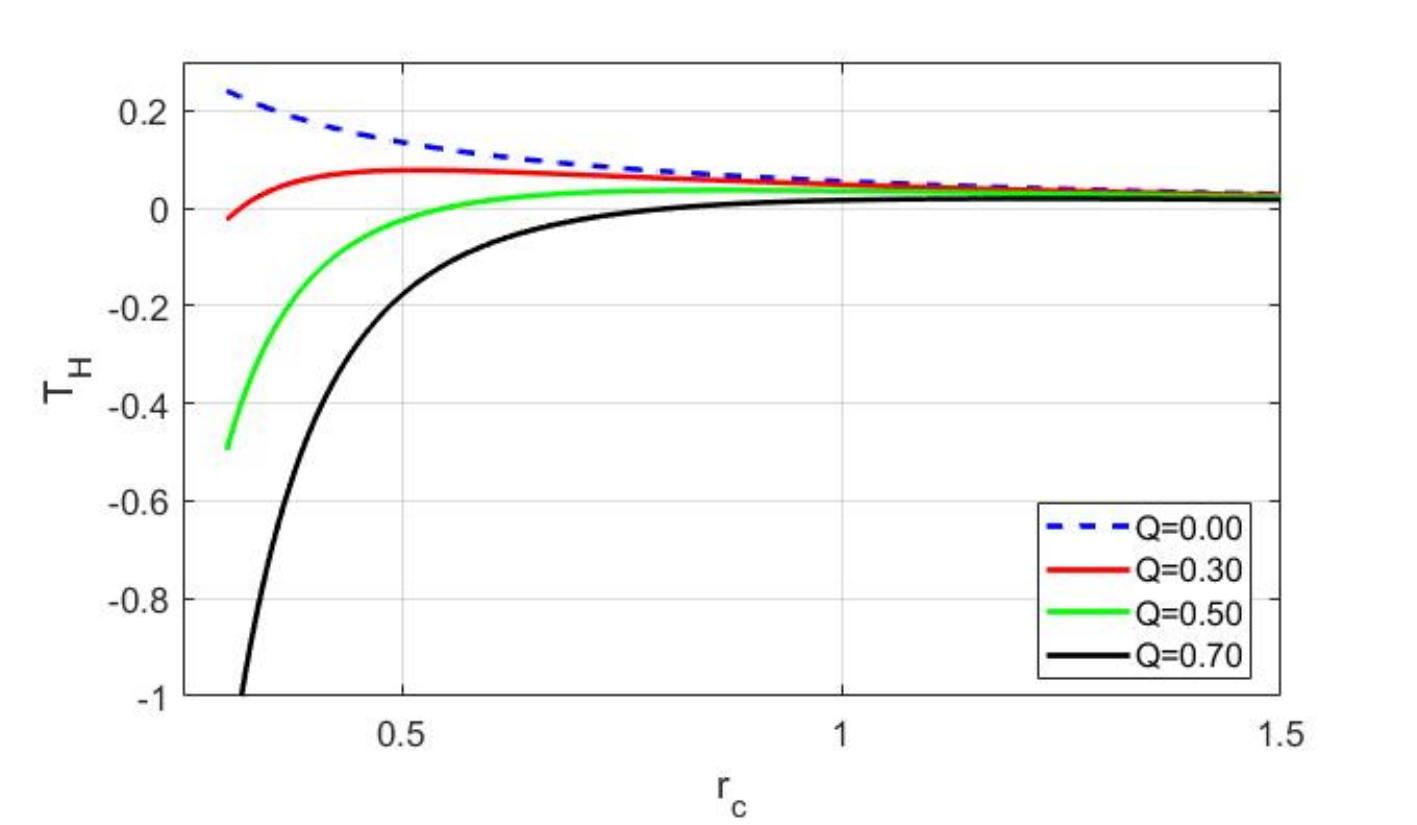}
	\caption{Hawking temperature  $T_H$ as a function of the horizon radius $r_c$ for different values of the charge $Q$.}
	\label{fig:T_H}
\end{figure}

Figure~6 illustrates the influence of the electric charge on the Hawking temperature of the Vaidya-Bonnor black hole. It is observed that increasing the electric charge leads to a systematic reduction of the Hawking temperature. From a physical viewpoint, the electric charge introduces an electromagnetic repulsive contribution that partially counteracts the gravitational attraction. This weakens the surface gravity at the event horizon and consequently decreases the thermal energy available for particle emission. As the charge approaches its critical value, the Hawking temperature tends toward lower values, indicating that the evaporation process becomes progressively less efficient. This behavior is consistent with the tendency of highly charged black holes to evaporate more slowly and remain thermodynamically more stable than weakly charged configurations. Therefore, the electric charge plays an important role in delaying black hole evaporation and contributes to the formation of a stable remnant when quantum effects dominate the final stages of the evaporation process.

Substituting Eqs. (41) and (40) into Eq. (39), the corrected entropy is written as

\begin{equation}
S_{c}=\pi r_{c}^{2} -\frac{\alpha}{2}\ln\left[\frac{1}{16\pi}\left(1-\frac{Q^{2}}{r_{c}^{2}}+3c\varepsilon r_{c}^{-3\varepsilon-1}\right)^{2}\right]
+\frac{\beta}{\pi r_{c}^{2}}.
\end{equation}

Equation (42) represents the complete second-order quantum corrected entropy including both the logarithmic correction and the inverse entropy correction. The first correction dominates for large black holes, whereas the second one becomes significant near the final stage of evaporation, leading to possible remnant formation.

\begin{figure}[htbp]
	\centering
	\includegraphics[width=0.75\textwidth]{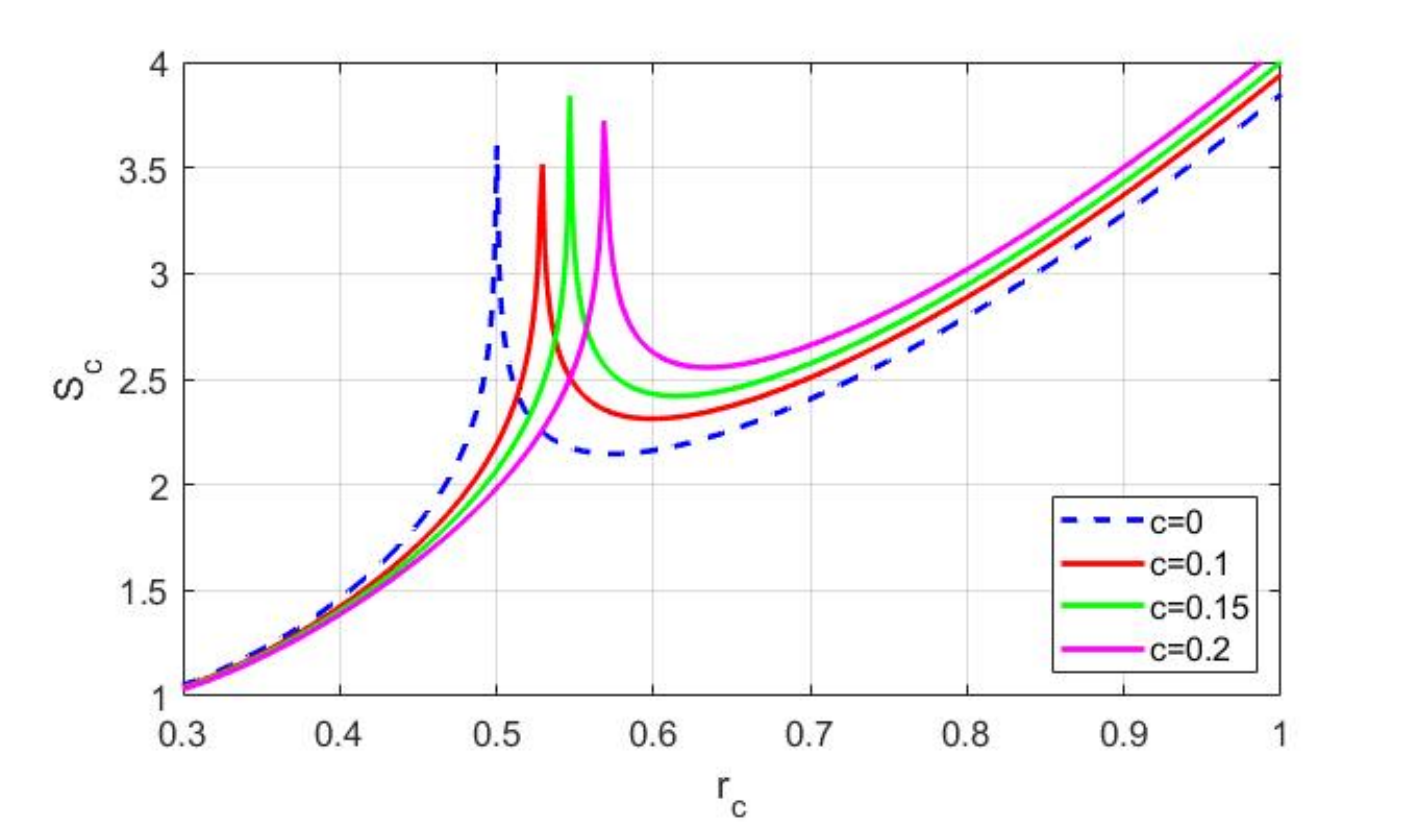}
	\caption{Corrected Entropy  $S_c$ as a function of the horizon radius $r_c$ for different values of the quintessence normalization parameter $c$.}
	\label{fig:S_c}
\end{figure}

Figure~7 illustrates the influence of the quintessence parameter $c$ on the corrected entropy. It is observed that the corrected entropy increases progressively with increasing values of $c$. Physically, the quintessence field behaves as a negative-pressure background that modifies the gravitational potential surrounding the Vaidya-Bonnor black hole. As the intensity of this dark-energy component increases, the evaporation process becomes less efficient due to the suppression of the Hawking temperature and the corresponding radiation rate. Consequently, quantum effects dominate at a relatively larger horizon radius, leading to the formation of a larger stable entropy. This result indicates that quintessence not only slows down the evaporation process but also determines the final size of the remnant, emphasizing the important role of the surrounding cosmological environment in the endpoint of black hole evaporation.
\bigskip

\begin{figure}[htbp]
	\centering
	\includegraphics[width=0.75\textwidth]{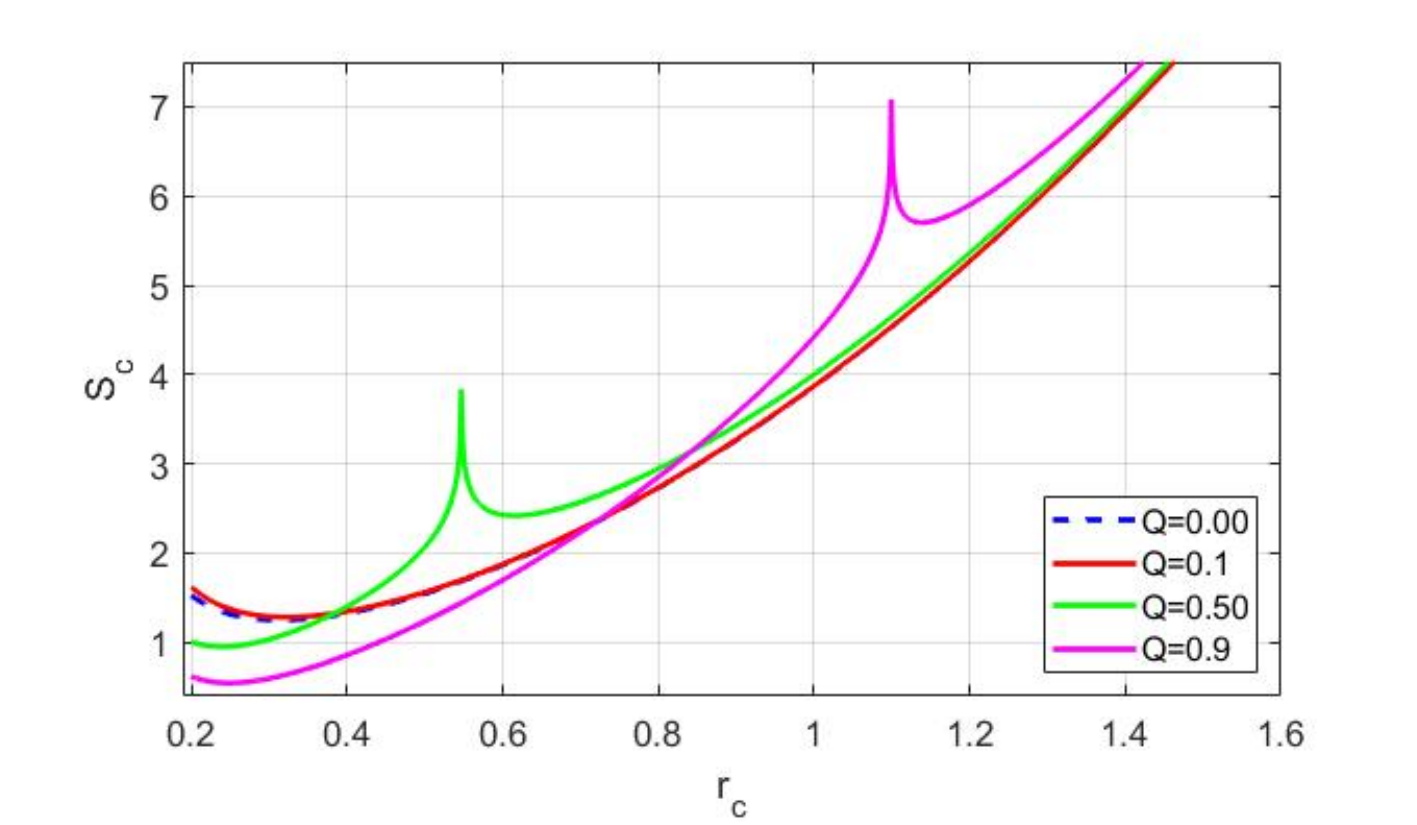}
	\caption{Corrected Entropy  $S_c$ as a function of the horizon radius $r_c$ for different values of the charge $Q$.}
	\label{fig:S_q}
\end{figure}

Figure~8 shows that the corrected entropy increases with the electric charge $Q$. The presence of electric charge generates an electromagnetic repulsive interaction that partially counteracts the gravitational attraction, thereby reducing the surface gravity at the event horizon. As a consequence, the Hawking temperature decreases and the evaporation process proceeds more slowly. Since the black hole loses its mass at a reduced rate, quantum corrections become significant before the horizon completely disappears, resulting in a larger corrected entropy. Therefore, highly charged black holes tend to preserve a larger residual horizon, highlighting the stabilizing influence of the electromagnetic field on the final stages of black hole evolution. \bigskip

\begin{figure}[htbp]
	\centering
	\includegraphics[width=0.75\textwidth]{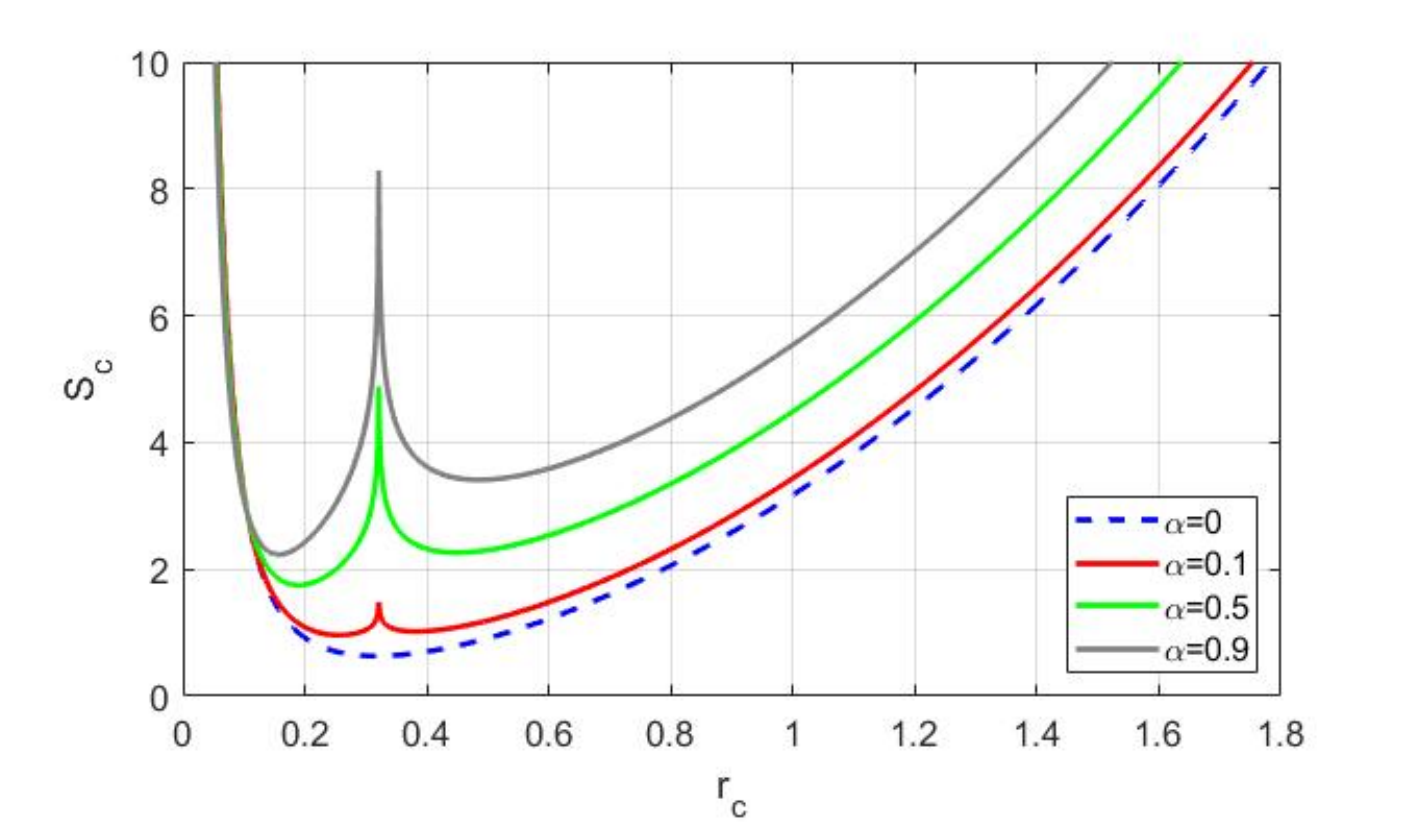}
	\caption{Corrected Entropy  $S_c$ as a function of the horizon radius $r_c$ for different values of the first order parameter $alpha$.}
	\label{fig:S_alp}
\end{figure}

Figure~9 demonstrates that the corrected entropy is sensitive to the first-order correction parameter $\alpha$. Increasing $\alpha$ produces a noticeable enlargement of the entropy, indicating that logarithmic quantum corrections become increasingly significant during the final stage of evaporation. Physically, these corrections originate from thermal fluctuations around the black hole equilibrium state and modify the thermodynamic quantities governing the evaporation process. The enlarged entropy implies that first-order quantum effects prevent the complete disappearance of the event horizon, thereby strengthening the stability of the final black hole configuration. This behavior confirms that logarithmic entropy corrections play an essential role in determining the thermodynamic endpoint of the evaporation process.

\bigskip

\begin{figure}[htbp]
	\centering
	\includegraphics[width=0.75\textwidth]{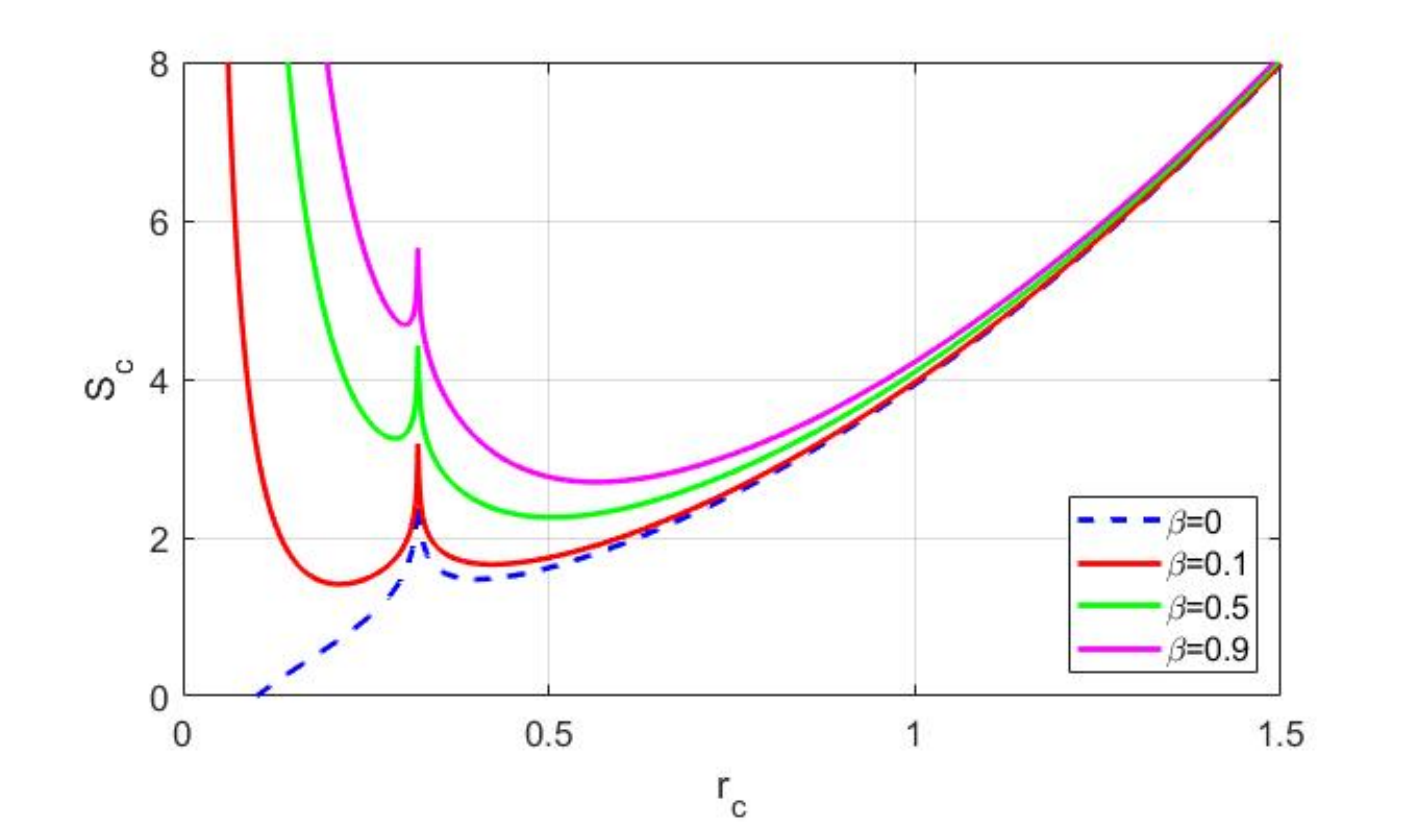}
	\caption{Corrected Entropy  $S_c$ as a function of the horizon radius $r_c$ for different values of the second parameter $bet$.}
	\label{fig:S_bet}
\end{figure}

Figure~10 presents the variation of the corrected entropy with the second-order correction parameter $\beta$. It is found that increasing $\beta$ further enlarges the entropy, indicating that higher-order quantum corrections become dominant in the final stage of black hole evaporation. Unlike the first-order logarithmic correction, the second-order contribution provides an additional stabilization mechanism by further suppressing the evaporation process near the Planck scale. Consequently, the black hole approaches a stable equilibrium configuration characterized by a finite horizon radius rather than undergoing complete evaporation. These results demonstrate that second-order quantum corrections considerably reinforce the formation and stability of black hole, highlighting their crucial role in describing the ultimate fate of Vaidya-Bonnor black holes surrounded by quintessence.

\subsection{Corrected Heat Capacity}

 In the canonical ensemble, the corrected heat capacity is defined by [51]

\begin{equation}
C_c=T_H\left(\frac{\partial S_c}{\partial T_H}\right)
=T_H\frac{\displaystyle\frac{\partial S_c}{\partial r_c}}
{\displaystyle\frac{\partial T_H}{\partial r_c}}.
\end{equation}

The derivative of the corrected entropy obtained previously,with respect to the horizon radius becomes

\begin{equation}
\frac{\partial S_c}{\partial r_c}=2\pi r_c+\frac{2\alpha}{r_c}+2\alpha\frac{T_H'}{T_H}-\frac{2\beta}{\pi r_c^3},
\end{equation}

where

\begin{equation}
T_H'
=\frac{1}{4\pi}\left(-\frac{1}{r_c^2}+\frac{3Q^2}{r_c^4}
-3c\varepsilon(2+3\varepsilon)r_c^{-3-3\varepsilon}\right).
\end{equation}

Substituting Eqs.~(41),(44) and (45) into Eq.~(43) and simplifying the resulting expression, the corrected heat capacity can be written as

\begin{equation}
C_c(r_c)=\frac{N(r_c)}{D(r_c)},
\end{equation}

where

\begin{equation}
N(r_c)= (2\pi^2 r_c^4 - 2\beta) \left(r_c^2-Q^2+3c \varepsilon r_c^{1-3\varepsilon}\right)-\alpha \pi r_c^2 \left(2Q^2-3c\varepsilon (3\varepsilon+1) r_c^{1-3\varepsilon}\right)
\end{equation}

and

\begin{equation}
D(r_c)={-\pi r_c [r_c^2-3Q^2+3c\varepsilon(3\varepsilon+2)r_c^{1-3\varepsilon}}
\end{equation}
For convenience, Eq.~(46) may also be written in the compact form

\begin{equation}
	\begin{aligned}
C_c(r_c)
=&
\frac{\left[(2\pi^2 r_c^4 - 2\beta)\left(r_c^2-Q^2+3c\varepsilon r_c^{1-3\varepsilon}\right)-\alpha \pi r_c^2\left(2Q^2-3 c\varepsilon(3\varepsilon+1)r_c^{1-3\varepsilon}\right)\right]}{-\pi r_c [r_c^2-3Q^2+3c\varepsilon(3\varepsilon+2)r_c^{1-3\varepsilon}]}
	\end{aligned}
\end{equation}

For $c=\alpha=\beta=0$, equation (49) reduced to classical heat capacity

\begin{equation}
C_Q=
\frac{2\pi r_c\left[r_c^{3\epsilon+3}-Q^2r_c^{3\epsilon}+3c\epsilon r_c\right]}{-r_c^{3\epsilon+1}+3Q^2r_c^{3\epsilon}-3c\epsilon(3\epsilon+2)r_c^2}.
\end{equation}

The thermal stability of the black hole is determined by the sign of the corrected heat capacity.

\begin{figure}[htbp]
	\centering
	\includegraphics[width=0.75\textwidth]{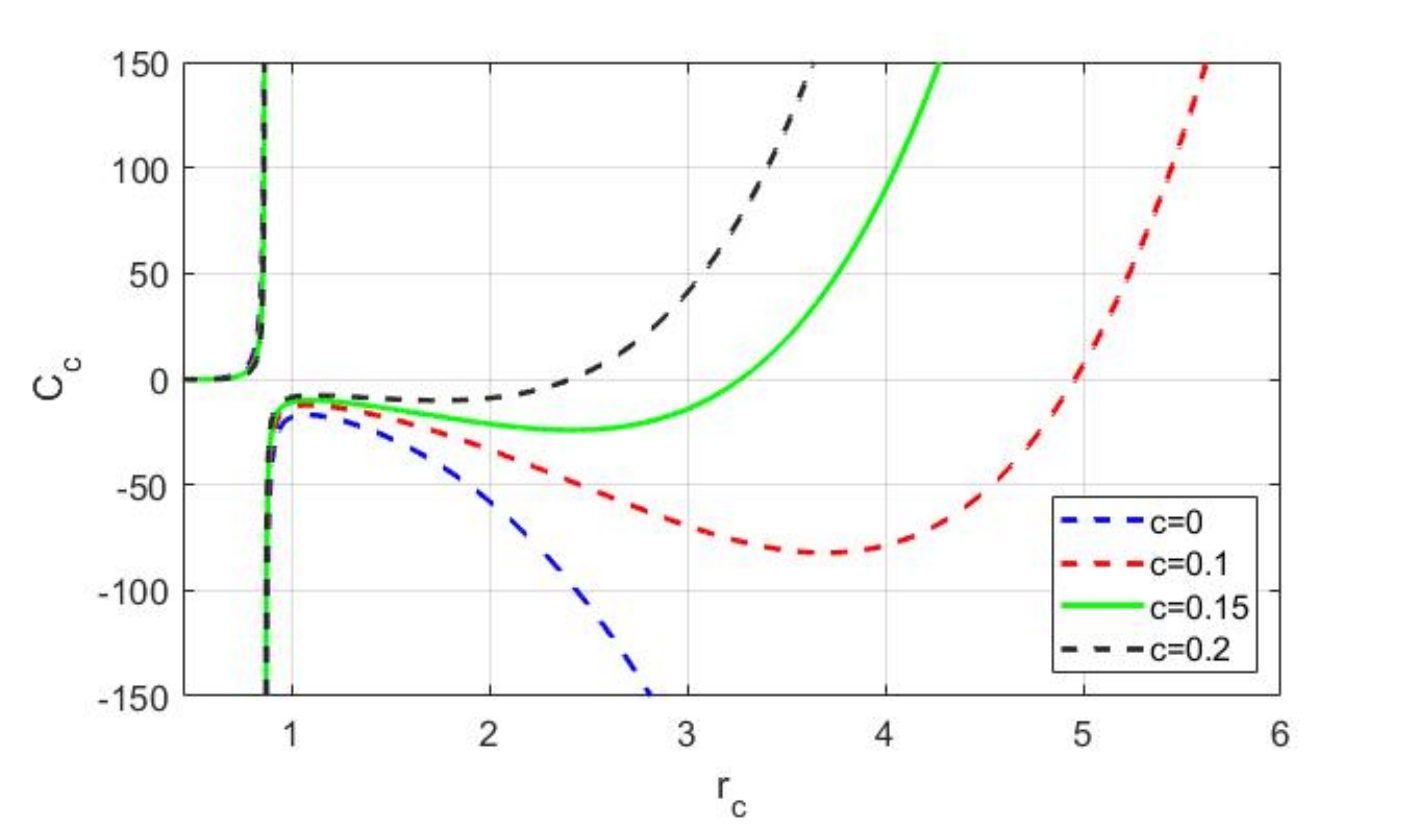}
	\caption{Corrected Heat Capacity  $C_c$ as a function of the horizon radius $r_c$ for different values of the quintessence normalization parameter $c$.}
	\label{fig:C_c}
\end{figure}

Figure~11 illustrates the effect of the quintessence parameter $c$ on the corrected heat capacity of the Vaidya-Bonnor black hole. It is observed that increasing $c$ significantly modifies the thermal response of the black hole and shifts the heat-capacity curves toward higher values of the event horizon radius. Physically, the negative-pressure background generated by the quintessence field alters the spacetime geometry and reduces the efficiency of the evaporation process. Consequently, the black hole is capable of storing a larger amount of thermal energy before undergoing substantial temperature variations. The change in the sign of the corrected heat capacity clearly distinguishes thermodynamically unstable and stable regions. The divergence of the heat capacity corresponds to a second-order phase transition, where the thermodynamic behavior changes abruptly. Therefore, a stronger quintessence field enhances the thermodynamic stability of the system and postpones the onset of instability, thereby supporting the persistence of a stable remnant during the final stage of evaporation.
\bigskip

\begin{figure}[htbp]
	\centering
	\includegraphics[width=0.75\textwidth]{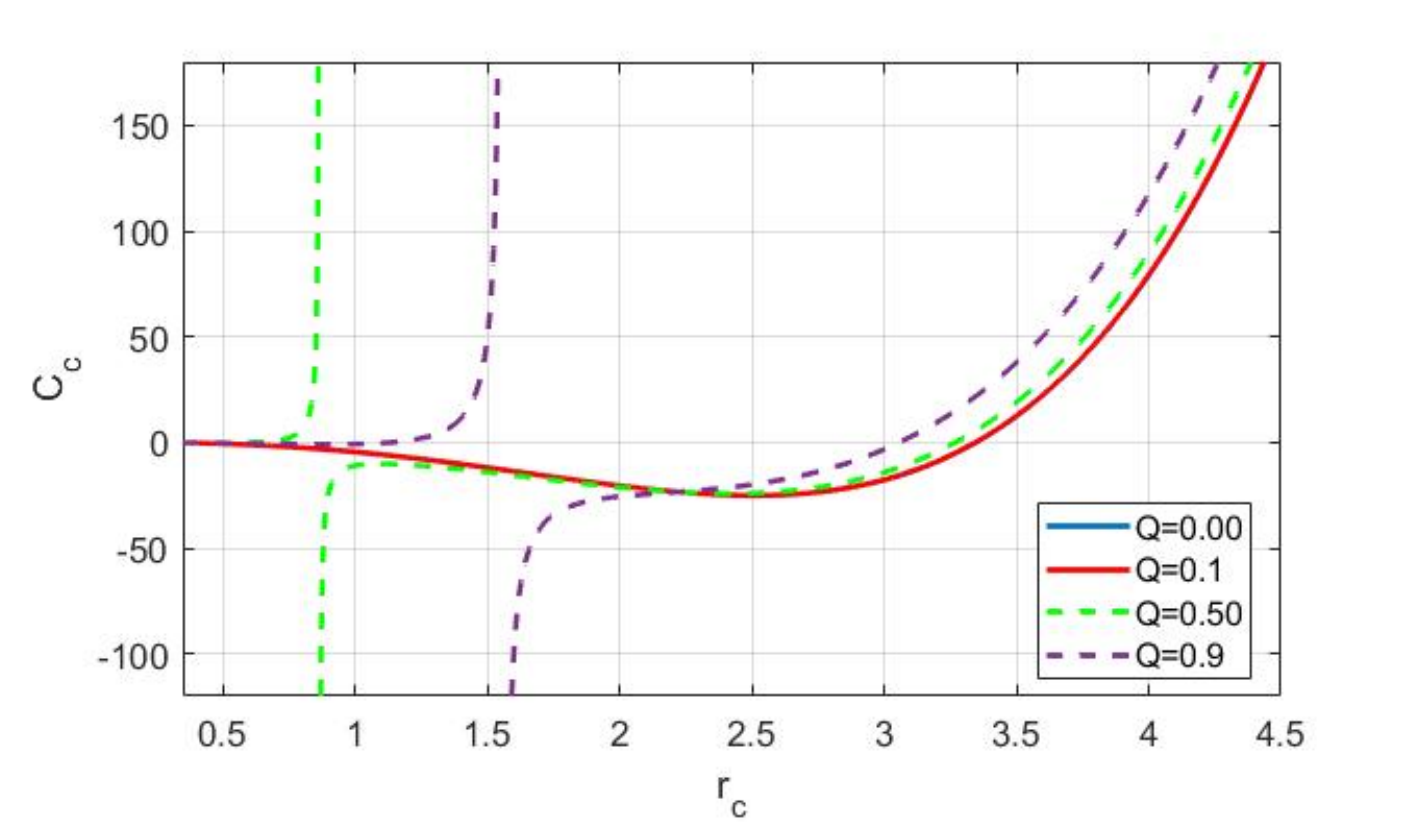}
	\caption{Corrected Heat Capacity  $C_c$ as a function of the horizon radius $r_c$ for different values of the charge $Q$.}
	\label{fig:C_q}
\end{figure}

Figure~12 presents the variation of the corrected heat capacity with the electric charge $Q$. The results indicate that increasing the electric charge modifies both the magnitude and the divergence point of the heat capacity. From a physical perspective, the electromagnetic repulsive interaction weakens the gravitational attraction, leading to a reduction in the Hawking temperature and consequently altering the thermal response of the black hole. The displacement of the divergence point demonstrates that the critical configuration associated with the second-order phase transition strongly depends on the electric charge. Highly charged black holes remain thermodynamically stable over a broader interval of the event horizon radius before reaching the critical point. This behavior confirms that the electric charge plays an essential role in regulating the thermal equilibrium of the system and contributes to delaying the final evaporation stage.

\bigskip

\begin{figure}[htbp]
	\centering
	\includegraphics[width=0.75\textwidth]{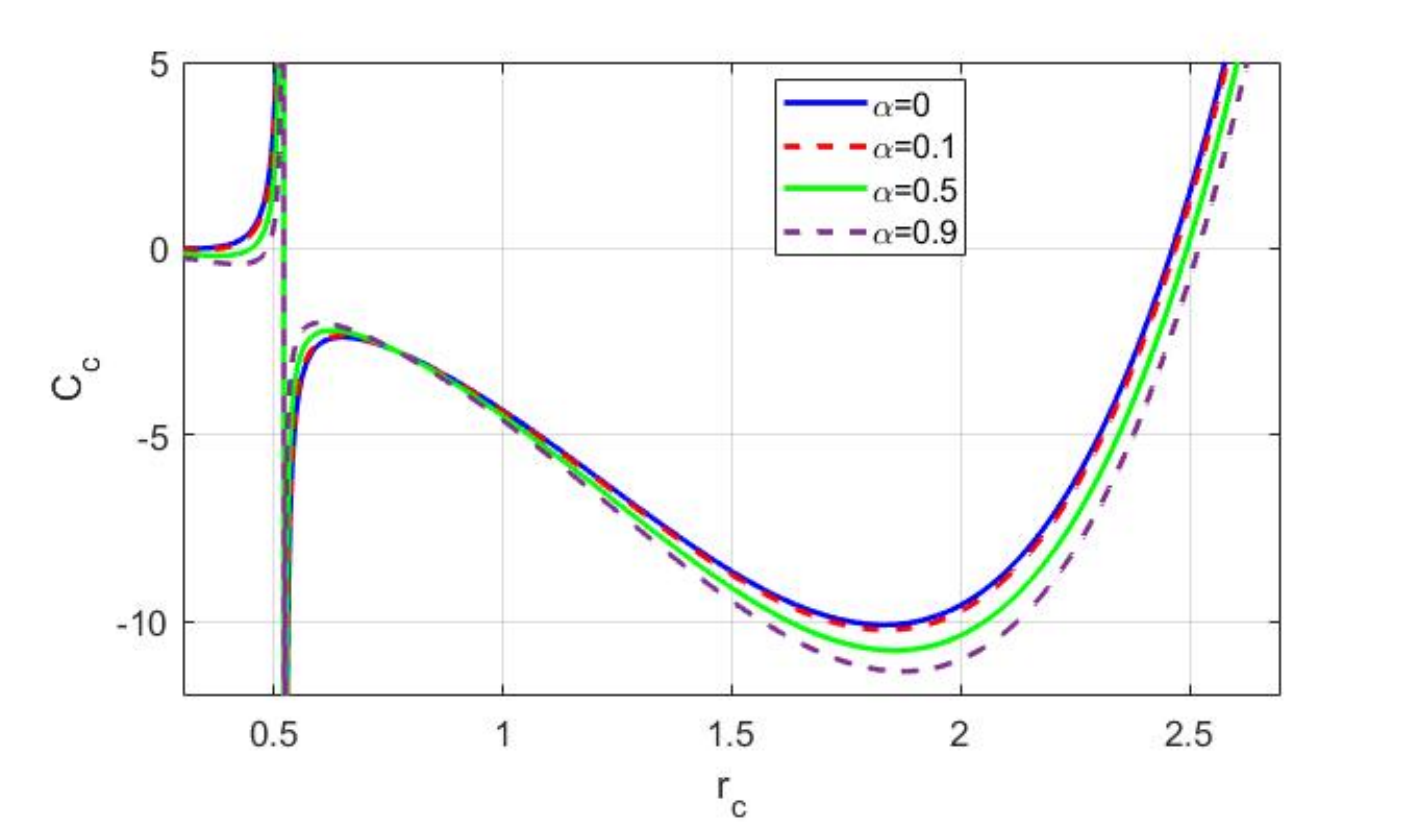}
	\caption{Corrected Heat Capacity  $C_c$ as a function of the horizon radius $r_c$ for different values of the first order parameter $\alpha$.}
	\label{fig:C_alp}
\end{figure}

Figure~13 shows that the first-order quantum correction parameter $\alpha$ has a significant influence on the corrected heat capacity. As $\alpha$ increases, the heat-capacity curves become progressively modified while preserving the overall thermodynamic structure of the system. Since the first-order correction originates from logarithmic entropy corrections induced by thermal fluctuations, it directly affects the exchange of heat between the black hole and its surroundings. The increase in $\alpha$ enhances the positive region of the heat capacity, indicating a wider domain of thermodynamic stability. Moreover, the position of the divergence point is slightly shifted, showing that quantum fluctuations influence the location of the phase transition without altering its second-order nature. These results demonstrate that first-order quantum corrections stabilize the thermodynamic evolution and delay the onset of thermal instability.
\bigskip

\begin{figure}[htbp]
	\centering
	\includegraphics[width=0.75\textwidth]{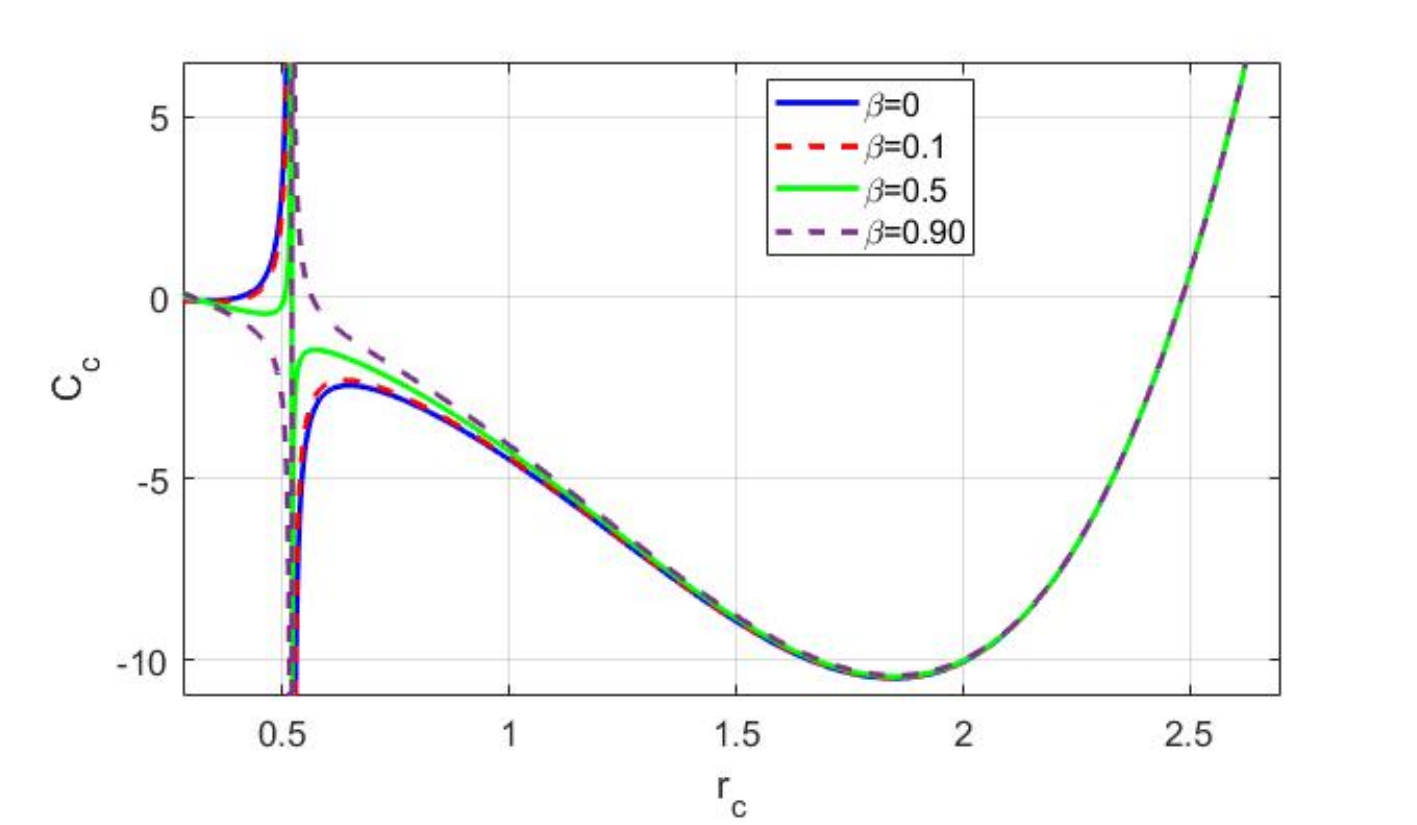}
	\caption{Corrected Heat Capacity  $C_c$ as a function of the horizon radius $r_c$ for different values of the second parameter $\beta$.}
	\label{fig:C_bet}
\end{figure}
Figure~14 illustrates the effect of the second-order quantum correction parameter $\beta$ on the corrected heat capacity. It is found that increasing $\beta$ produces a more pronounced modification of the heat-capacity curves than the first-order correction. Physically, second-order quantum fluctuations become increasingly important as the black hole approaches the final stage of evaporation, where semiclassical approximations are no longer sufficient. The enlargement of the stable region together with the displacement of the divergence point indicates that higher-order quantum effects considerably improve the thermodynamic stability of the black hole. Although the second-order phase transition remains preserved, its critical configuration is significantly influenced by the magnitude of $\beta$. These findings confirm that second-order quantum corrections play a fundamental role in determining the thermodynamic behavior of the black hole and reinforce the formation of a stable remnant after the evaporation process has effectively ceased.

\subsection{Corrected Enthalpy, Corrected Mass and Corrected Internal Energy}

The corrected enthalpy$(H_{c}=M_{c})$ is obtained from the first law of black hole thermodynamics. Assuming that the thermodynamic pressure remains constant, the differential form reduces to [55]

\begin{equation}
dH_{c}=T_{H}dS_{c},
\end{equation}

which leads to

\begin{equation}
H_{c}=\int T_{H} dS_{c}.
\end{equation}

Using the corrected entropy Eq. (42), its differential with respect to the horizon radius is

\begin{equation}
dS_{c}=\left[2\pi r_{c}-\frac{\alpha}{r_{c}}-\frac{\alpha}{T_{H}}\frac{dT_{H}}{dr_{c}}-\frac{2\beta}{\pi r_{c}^{3}}\right]dr_{c}.
\end{equation}

Multiplying Eq. (53) by the Hawking temperature gives

\begin{equation}
T_{H}dS_{c}
=\left[2\pi r_{c}T_{H}-\frac{\alpha T_{H}}{r_{c}}-\alpha\frac{dT_{H}}{dr_{c}}-\frac{2\beta T_{H}}{\pi r_{c}^{3}}\right]dr_{c}.
\end{equation}

Consequently,

\begin{equation}
H_{c}=2\pi\int r_{c}T_{H}dr_{c}-\alpha\int\frac{T_{H}}{r_{c}}dr_{c}-\alpha T_{H}-\frac{2\beta}{\pi}\int\frac{T_{H}}{r_{c}^{3}}dr_{c}.
\end{equation}

Substituting the Hawking temperature Eq. (41) into Eq. (55), each integral can be evaluated analytically. After straightforward calculations, one finally obtains

\begin{equation}
	\begin{aligned}
	H_{c}
	=&
	\frac12 \left(r_{c}+\frac{Q^{2}}{r_{c}}-cr_{c}^{-3\varepsilon}\right)+\frac{\alpha Q^{2}}{6\pi r_{c}^{3}}-\frac{\alpha c\varepsilon^{2}}{\pi(\varepsilon+1)}r_{c}^{-3\varepsilon-3}
	\\
	&
	+\frac{\beta}{6\pi^{2}r_{c}^{3}}-\frac{\beta Q^{2}}{10\pi^{2}r_{c}^{5}}+\frac{3\beta c\varepsilon}{2\pi^{2}(3\varepsilon+4)}r_{c}^{-3\varepsilon-4}.
	\end{aligned}
\end{equation}

 The correction proportional to $\alpha$ becomes important for intermediate black holes, whereas the $\beta$-dependent contribution dominates near the final stage of black hole evaporation. Compared with the classical case, quantum corrections become dominant for small black holes.

\begin{figure}[htbp]
	\centering
	\includegraphics[width=0.75\textwidth]{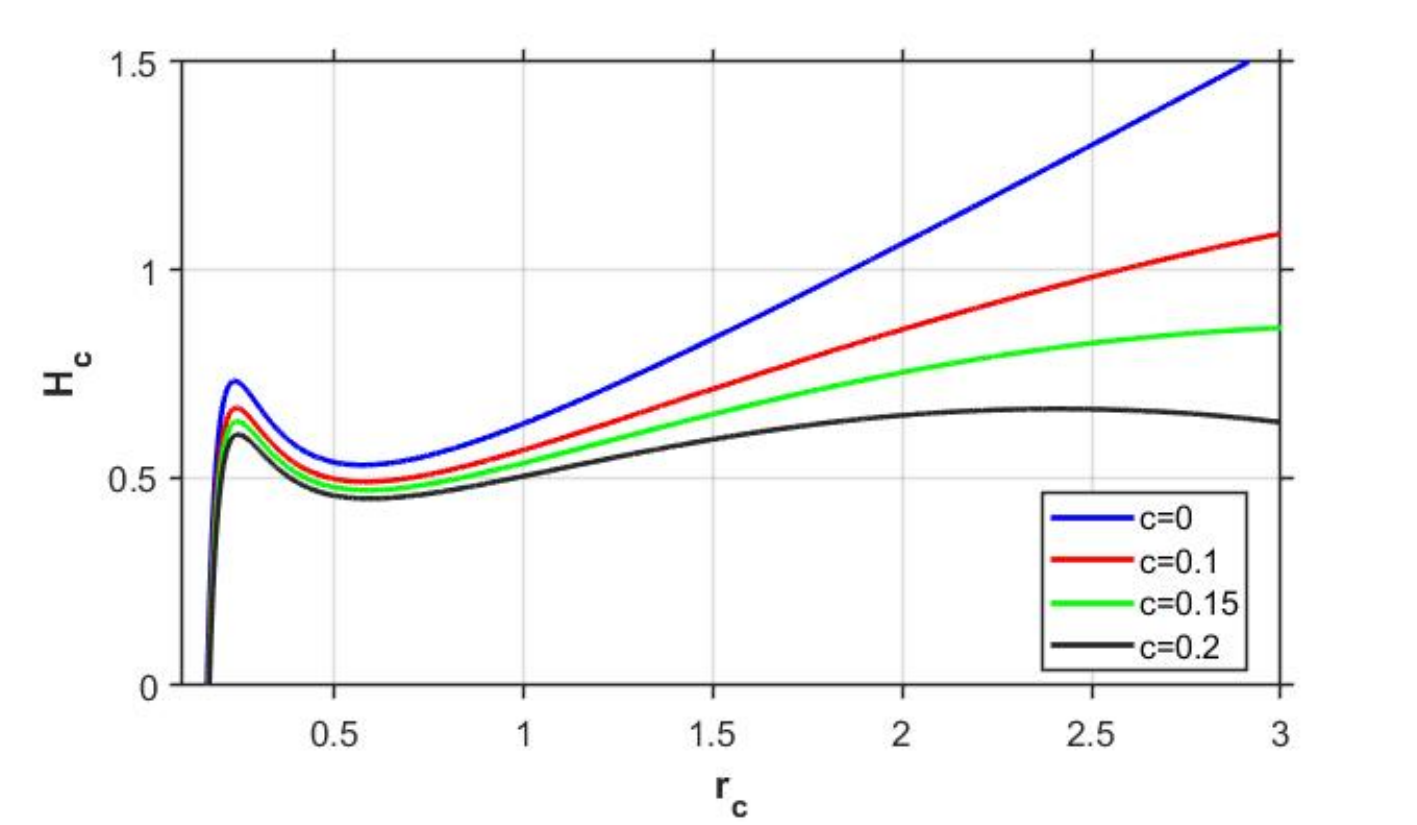}
	\caption{Corrected Enthalpy  $H_c$ as a function of the horizon radius $r_c$ for different values of the quintessence normalization parameter $c$.}
	\label{fig:H_c}
\end{figure}
Figure~15 illustrates the influence of the quintessence parameter $c$ on the corrected enthalpy of the Vaidya-Bonnor black hole. It is observed that the corrected enthalpy decreases progressively as the quintessence parameter increases over the entire range of the event horizon radius. Physically, the negative-pressure background associated with quintessence modifies the spacetime geometry and reduces the effective gravitational energy stored by the black hole. Since the enthalpy represents the total energy content of the gravitational system, including the work required to sustain the surrounding spacetime, the presence of a stronger quintessence field lowers the total thermodynamic energy available. Consequently, the black hole evolves toward a less energetic but more thermodynamically stable configuration. This behavior further indicates that the surrounding dark-energy field actively participates in regulating the energy balance of the black hole and favors the persistence of a stable remnant during the late stages of evaporation.
\bigskip

\begin{figure}[htbp]
	\centering
	\includegraphics[width=0.75\textwidth]{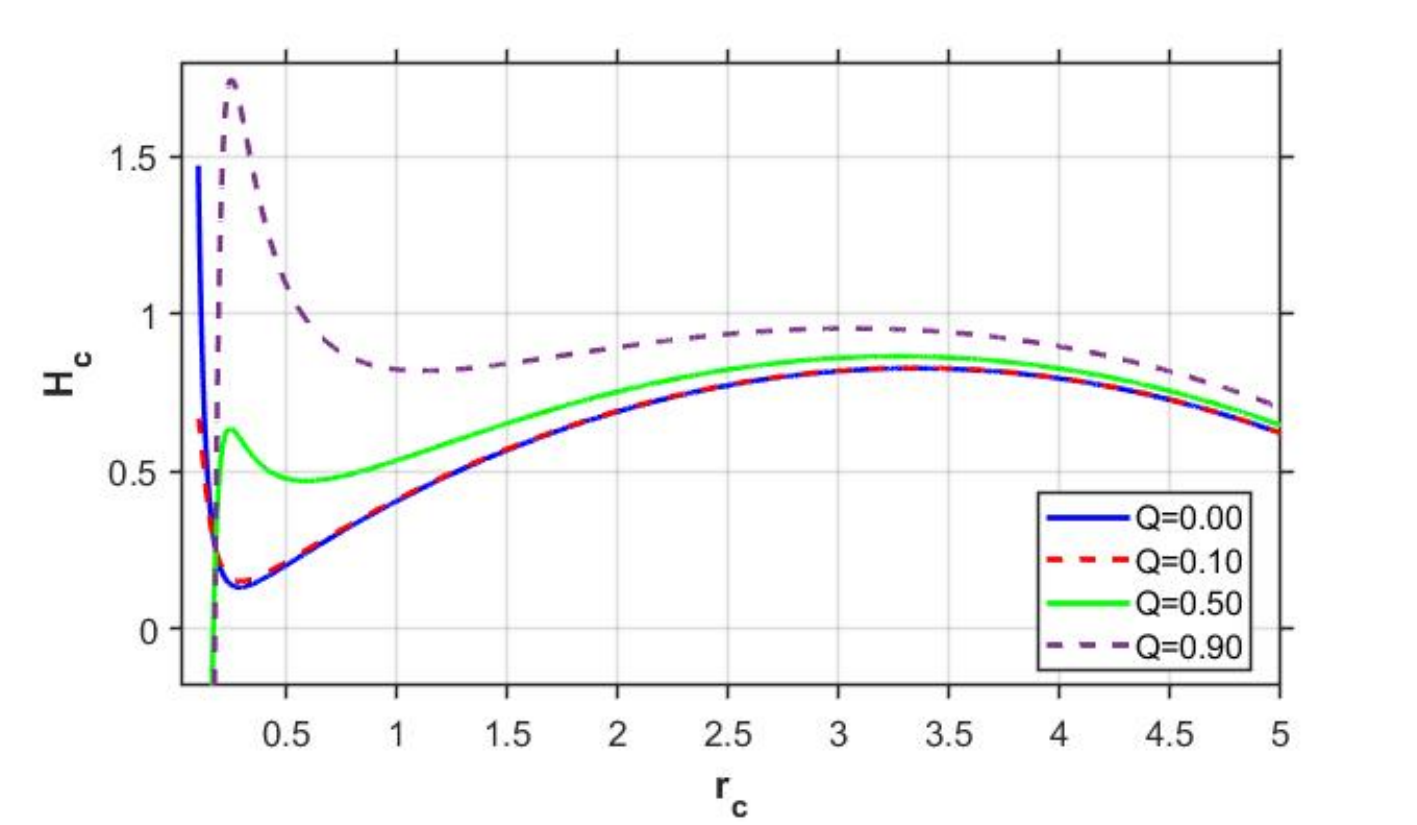}
	\caption{Corrected Enthalpy  $H_c$  as a function of the horizon radius $r_c$ for different values of the charge $Q$.}
	\label{fig:H_q}
\end{figure}

Figure~16 presents the variation of the corrected enthalpy with the electric charge $Q$. The results show that increasing the electric charge enhances the corrected enthalpy throughout the evolution of the event horizon. From a physical viewpoint, the electric charge contributes additional electromagnetic energy to the gravitational system, thereby increasing its total energy content. Although the charge suppresses the Hawking temperature and slows down the evaporation process, it simultaneously stores a larger amount of energy within the black hole configuration. Consequently, highly charged black holes possess larger corrected enthalpy and require more energy to reach the final evaporation stage. This result demonstrates that the electromagnetic field plays an essential role in determining the global thermodynamic energy and contributes significantly to the stabilization of the remnant configuration.
\bigskip

\begin{figure}[htbp]
	\centering
	\includegraphics[width=0.75\textwidth]{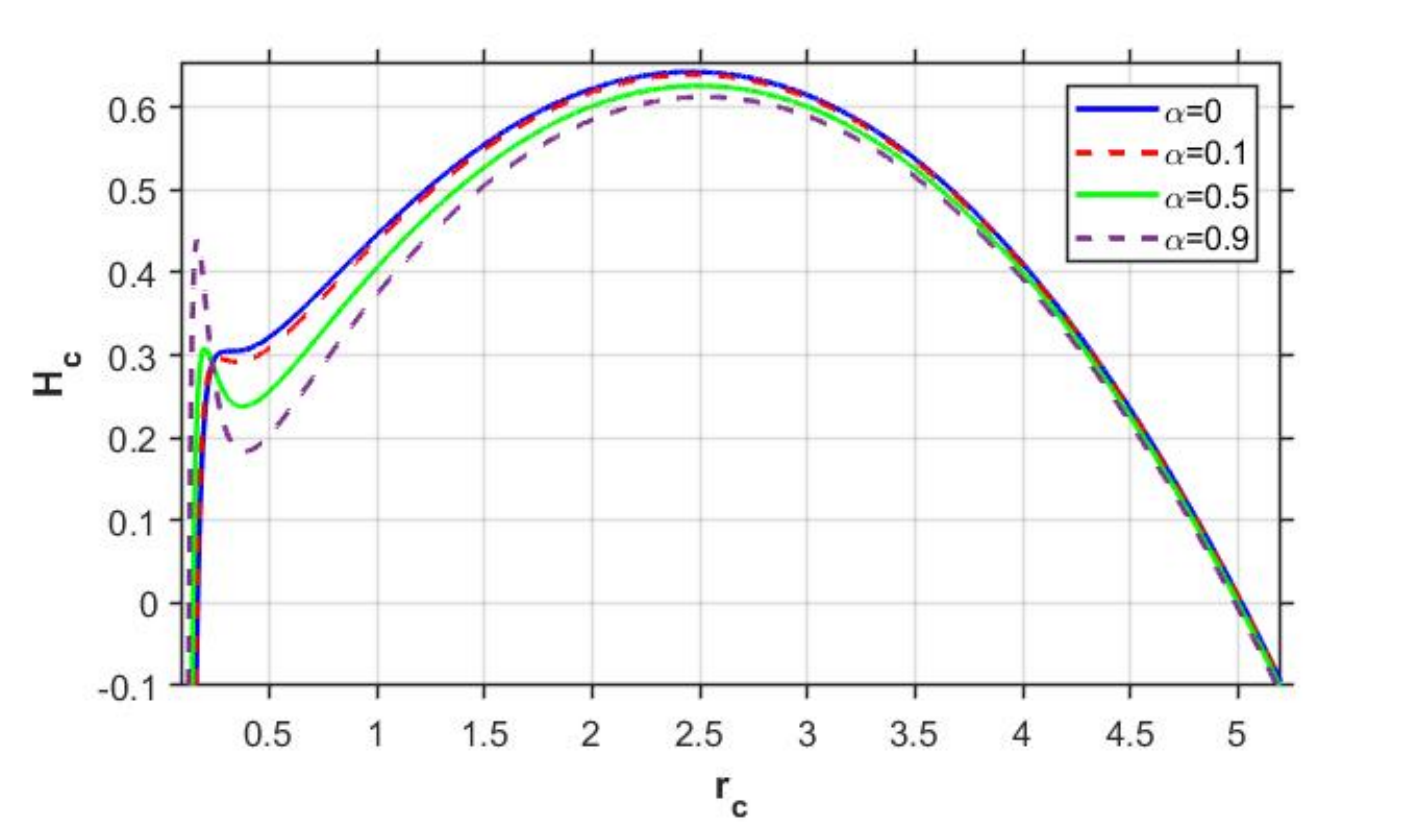}
	\caption{Corrected Enthalpy  $H_c$ as a function of the horizon radius $r_c$ for different values of the first order parameter $alpha$.}
	\label{fig:H_alp}
\end{figure}
Figure~17 shows the effect of the first-order correction parameter $\alpha$ on the corrected enthalpy. As $\alpha$ increases, noticeable modifications appear in the enthalpy curves, reflecting the contribution of logarithmic quantum corrections to the total thermodynamic energy. These corrections originate from thermal fluctuations around the equilibrium state and become increasingly important as the black hole approaches the final stage of evaporation. The corresponding variation of the corrected enthalpy indicates that first-order quantum effects alter the energy exchange between the black hole and its surroundings while preserving the overall thermodynamic behavior. Consequently, logarithmic corrections contribute to maintaining the stability of the black hole and delay the complete release of its internal energy.
\bigskip

\begin{figure}[htbp]
	\centering
	\includegraphics[width=0.75\textwidth]{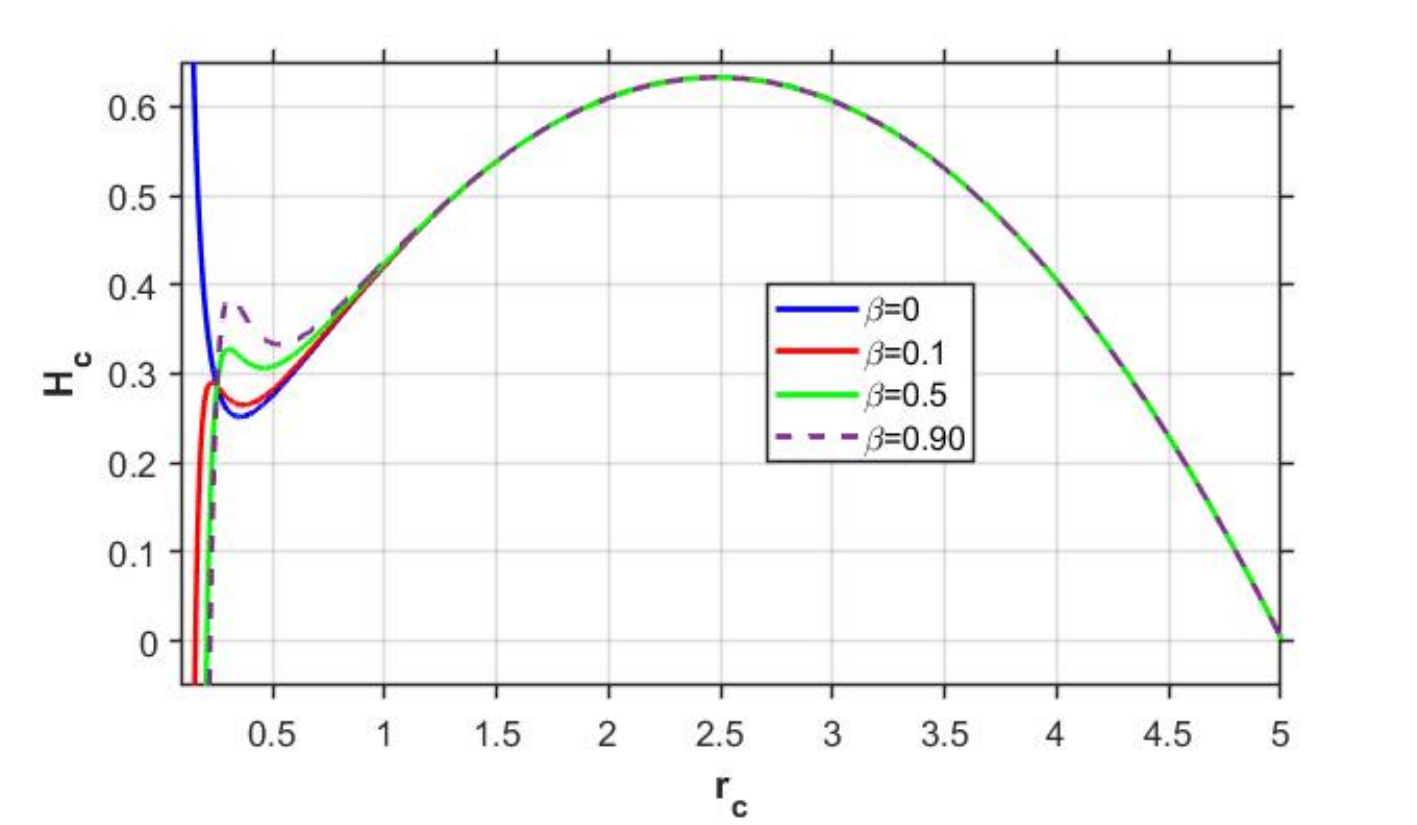}
	\caption{Corrected Enthalpy  $H_c$ as a function of the horizon radius $r_c$ for different values of the second parameter $\beta$.}
	\label{fig:H_bet}
\end{figure}
Figure~18 illustrates the dependence of the corrected enthalpy on the second-order correction parameter $\beta$. It is found that increasing $\beta$ produces a more pronounced modification of the corrected enthalpy than the first-order correction, emphasizing the growing importance of higher-order quantum effects near the endpoint of evaporation. Physically, second-order quantum corrections further modify the distribution of thermodynamic energy and reinforce the deviation from the semiclassical description. The progressive variation of the corrected enthalpy demonstrates that higher-order fluctuations significantly influence the energetic properties of the black hole while preserving its thermodynamic consistency. These results indicate that second-order quantum corrections constitute an essential ingredient for describing the complete thermodynamic evolution of Vaidya-Bonnor black holes surrounded by quintessence and provide additional support for the formation of a stable black hole remnant.

\subsection{Corrected Helmholtz and Corrected Gibbs Free Energy}

The corrected Helmholtz free energy is obtained from the thermodynamic definition [55]

\begin{equation}
F_{c}=-\int S_{c}dT_{H}.
\end{equation}

Applying integration by parts,

\begin{equation}
\int S_{c}dT_{H} = S_{c}T_{H}-\int T_{H}dS_{c},
\end{equation}

thus,

\begin{equation}
F_{c}=-S_{c}T_{H}+\int T_{H}dS_{c}.
\end{equation}

From the previous subsection,

\begin{equation}
\int T_{H}dS_{c}=H_{c},
\end{equation}

therefore

\begin{equation}
F_{c}=H_{c}-T_{H}S_{c}.
\end{equation}

Consequently,

\begin{equation}
\begin{aligned}
T_{H}S_{c}
=&
\frac{r_{c}}{4}\left(1-\frac{Q^{2}}{r_{c}^{2}}+3c\varepsilon r_{c}^{-3\varepsilon-1}\right)-\frac{\alpha}{8\pi r_{c}}
\left(1-\frac{Q^{2}}{r_{c}^{2}}+3c\varepsilon r_{c}^{-3\varepsilon-1}\right)
\\
&
\times \ln\left[\frac{1}{16\pi}\left(1-\frac{Q^{2}}{r_{c}^{2}}+3c\varepsilon r_{c}^{-3\varepsilon-1}\right)^{2}\right]+\frac{\beta}{4\pi^{2}r_{c}^{3}}\left(1-\frac{Q^{2}}{r_{c}^{2}}+3c\varepsilon r_{c}^{-3\varepsilon-1}\right).
\end{aligned}
\end{equation}

Substituting Eqs. (56) and (62) into Eq. (61), the corrected Helmholtz free energy becomes

\begin{equation}
	\begin{aligned}
F_{c}
	=&
\frac12\left(r_{c}+\frac{Q^{2}}{r_{c}}-cr_{c}^{-3\varepsilon}\right)+\frac{\alpha Q^{2}}{6\pi r_{c}^{3}}-\frac{\alpha c\varepsilon^{2}}{\pi(\varepsilon+1)}r_{c}^{-3\varepsilon-3}+\frac{\beta}{6\pi^{2}r_{c}^{3}}
	\\
	&
-\frac{\beta Q^{2}}{10\pi^{2}r_{c}^{5}}+\frac{3\beta c\varepsilon}{2\pi^{2}(3\varepsilon+4)}r_{c}^{-3\varepsilon-4}-\frac{r_{c}}{4}\left(1-\frac{Q^{2}}{r_{c}^{2}}+3c\varepsilon r_{c}^{-3\varepsilon-1}\right)
	\\
	&
+\frac{\alpha}{8\pi r_{c}}\left(1-\frac{Q^{2}}{r_{c}^{2}}+3c\varepsilon r_{c}^{-3\varepsilon-1}\right)\ln\left[\frac{1}{16\pi}\left(1-\frac{Q^{2}}{r_{c}^{2}}+3c\varepsilon r_{c}^{-3\varepsilon-1}\right)^{2}\right]
	\\
	&
-\frac{\beta}{4\pi^{2}r_{c}^{3}}\left(1-\frac{Q^{2}}{r_{c}^{2}}+3c\varepsilon r_{c}^{-3\varepsilon-1}\right).
	\end{aligned}
\end{equation}

Equation (63) shows that the corrected enthalpy reduces to the classical black-hole mass in the absence of thermal fluctuations, namely for $\alpha=\beta=0$. The logarithmic correction mainly affects intermediate horizon radii, whereas the inverse-entropy correction becomes dominant during the final stage of the evaporation process.

\begin{figure}[htbp]
	\centering
	\includegraphics[width=0.75\textwidth]{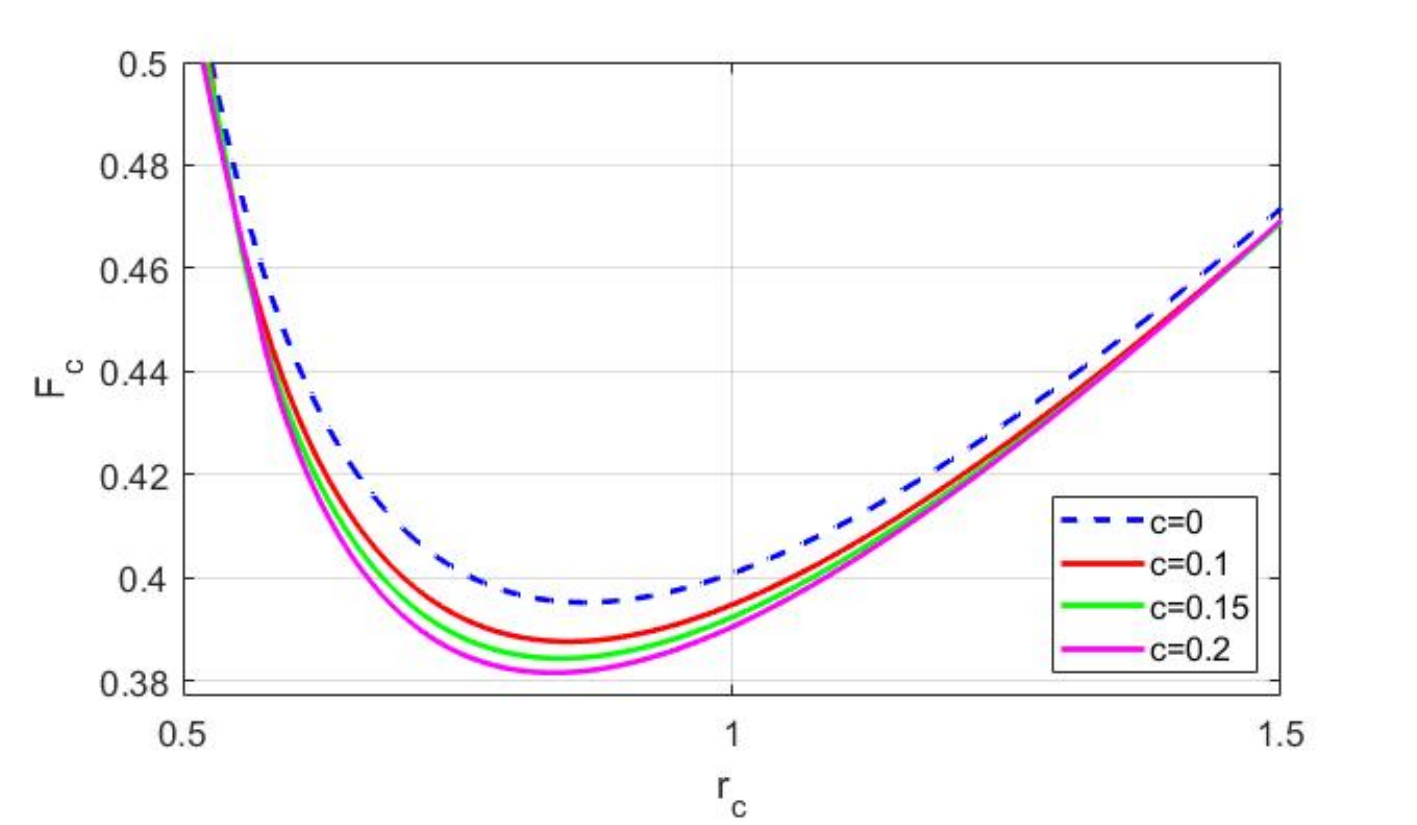}
	\caption{Corrected Helmholtz $F_c$  as a function of the horizon radius $r_c$ for different values of the quintessence normalization parameter $c$.}
	\label{fig:F_c}
\end{figure}
Figure~19 illustrates the influence of the quintessence parameter $c$ on the corrected Helmholtz free energy of the Vaidya-Bonnor black hole. It is observed that increasing the quintessence parameter significantly modifies the Helmholtz free energy over the entire range of the event horizon radius. Physically, the negative-pressure background generated by the quintessence field alters the balance between the internal energy and thermal energy of the black hole. Since the Helmholtz free energy determines the amount of useful energy available to perform thermodynamic work at constant temperature, its variation reflects the progressive redistribution of energy induced by the surrounding dark-energy field. The corresponding behavior indicates that stronger quintessence drives the black hole toward a more thermodynamically favorable equilibrium configuration. Consequently, the evaporation process becomes less efficient, while the stability of the final remnant configuration is enhanced by the presence of the quintessence field.
\bigskip

\begin{figure}[htbp]
	\centering
	\includegraphics[width=0.75\textwidth]{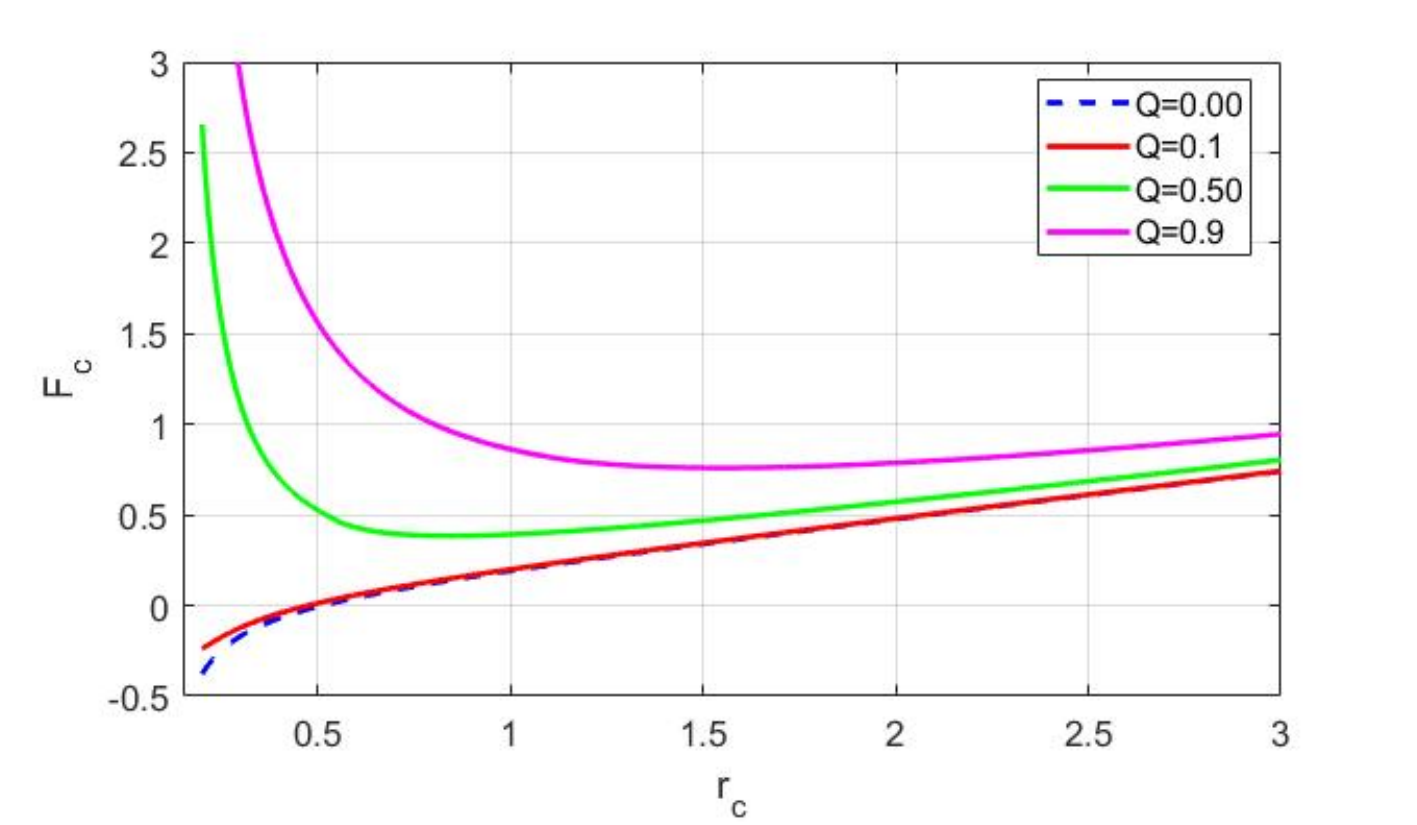}
	\caption{Corrected Helmholtz  $F_c$ as a function of the horizon radius $r_c$ for different values of the charge $Q$. The remaining parameters are kept fixed.}
	\label{fig:F_q}
\end{figure}

Figure~20 presents the dependence of the corrected Helmholtz free energy on the electric charge $Q$. The results reveal that the electric charge substantially modifies the free-energy profile throughout the evolution of the event horizon. From a physical viewpoint, the electromagnetic field contributes additional energy to the black hole while simultaneously reducing the Hawking temperature through the weakening of the surface gravity. The competition between these two effects changes the thermodynamic equilibrium of the system and modifies the amount of available free energy. As the electric charge increases, the black hole approaches a more stable thermodynamic configuration, requiring a larger amount of energy to complete the evaporation process. This behavior confirms that the electromagnetic interaction plays an important role in regulating the energetic stability of charged Vaidya-Bonnor black holes.\bigskip

\begin{figure}[htbp]
	\centering
	\includegraphics[width=0.75\textwidth]{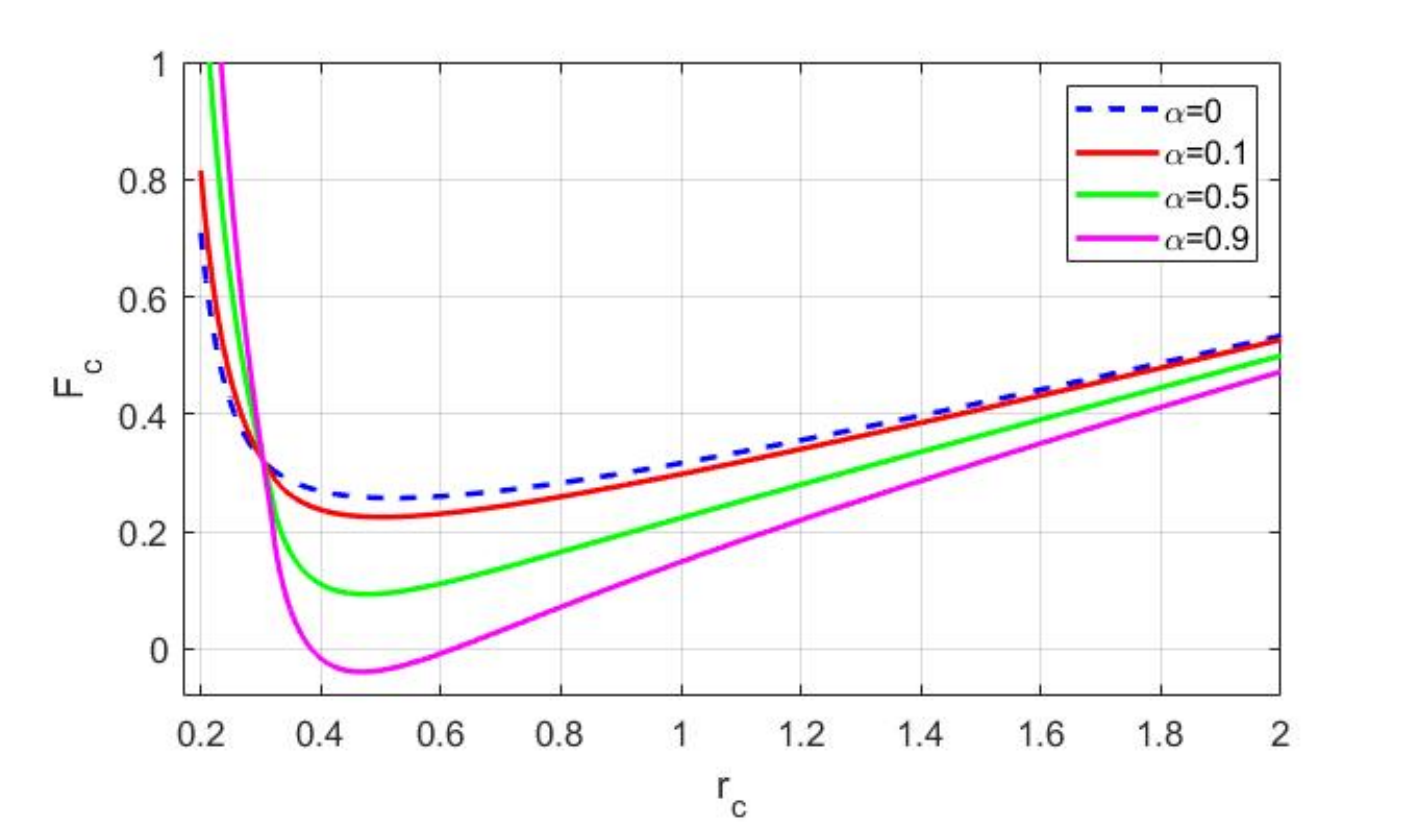}
	\caption{Corrected Helmholtz  $F_c$ as a function of the horizon radius $r_c$ for different values of the first order parameter $alpha$.}
	\label{fig:F_alp}
\end{figure}

Figure~21 shows the effect of the first-order correction parameter $\alpha$ on the corrected Helmholtz free energy. As $\alpha$ increases, noticeable variations appear in the free-energy curves, demonstrating the contribution of logarithmic quantum corrections to the thermodynamic equilibrium of the black hole. These corrections originate from thermal fluctuations around the equilibrium state and become increasingly relevant during the late stages of evaporation. Consequently, the available free energy is redistributed by quantum effects, modifying the thermodynamic evolution without changing the overall physical consistency of the system. The corresponding behavior indicates that first-order quantum corrections improve the thermodynamic stability of the black hole and contribute to delaying the complete evaporation process.
\bigskip

\begin{figure}[htbp]
	\centering
	\includegraphics[width=0.75\textwidth]{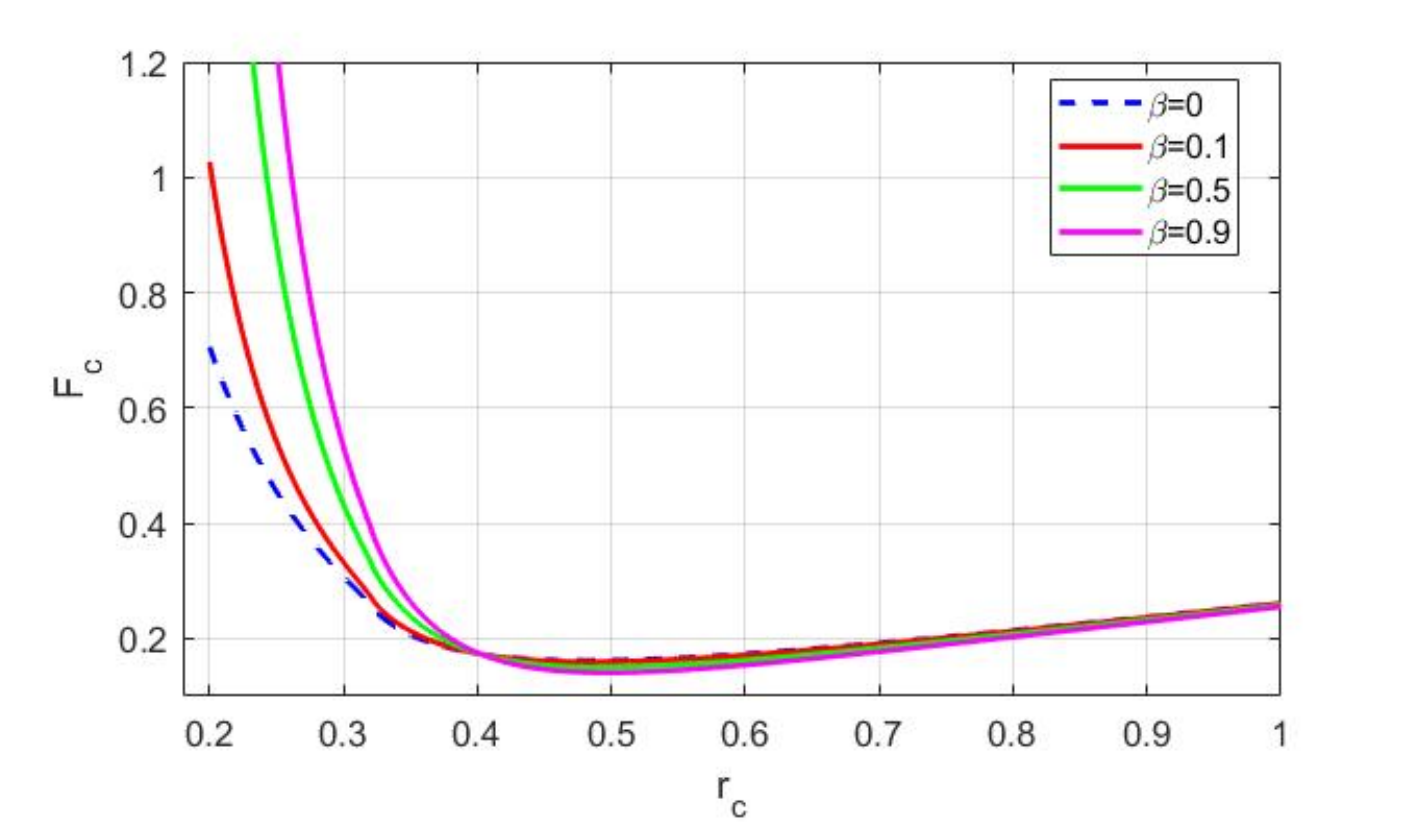}
	\caption{Corrected Helmholtz $F_c$ as a function of the horizon radius $r_c$ for different values of the second parameter $\beta$.}
	\label{fig:F_bet}
\end{figure}

Figure~22 illustrates the influence of the second-order correction parameter $\beta$ on the corrected Helmholtz free energy. It is found that increasing $\beta$ produces a more pronounced modification of the free-energy profile than the first-order correction, highlighting the increasing importance of higher-order quantum fluctuations near the endpoint of evaporation. Physically, second-order quantum corrections further modify the thermodynamic equilibrium by reducing the amount of free energy available for the evaporation process. As a consequence, the black hole evolves toward a more stable final configuration characterized by finite thermodynamic quantities. These results demonstrate that higher-order quantum effects play a crucial role in governing the final stages of black hole evaporation and provide additional evidence supporting the formation and long-term stability of black hole remnants surrounded by quintessence.

The corrected Gibbs free energy is defined as

\begin{equation}
G_{c}=M_{c}-T_{H}S_{c},
\end{equation}

Since, the corrected enthalpy coincides with the corrected mass,

\begin{equation}
G_{c}=H_{c}-T_{H}S_{c}
\end{equation}
The classical contribution is modified by the logarithmic correction parameter $\alpha$ and the inverse entropy correction parameter $\beta$. The behavior of the Gibbs free energy provides valuable information about the thermodynamic stability and the occurrence of possible phase transitions. In particular, the sign change of the Gibbs free energy distinguishes thermodynamically preferred configurations from unstable ones, whereas its extrema are closely related to the critical behavior of the black hole system.

\section{Analytical Determination of Quantum Black Hole Remnants}

One of the most remarkable consequences of quantum thermal fluctuations is the possible formation of a stable black hole remnant at the endpoint of the evaporation process. In the semiclassical description, the continuous emission of Hawking radiation may eventually drive the black hole to complete evaporation. However, the presence of logarithmic and higher-order thermal fluctuation corrections can significantly modify the thermodynamic behavior near the final stage of evaporation, leading to the existence of a finite remnant characterized by a non-vanishing horizon radius and residual mass [41-63].

In this section, we derive analytical expressions for the corrected remnant radius and the corresponding remnant mass of a Vaidya-Bonnor black hole surrounded by quintessence. Following the thermodynamic remnant criterion adopted in this work, we identify the remnant configuration with the zero of the corrected heat capacity,

\begin{equation}
C_c=0.
\end{equation}

Using the analytical expression obtained in the previous section, the above condition is equivalent to

\begin{equation}
N(r_c)=0,
\end{equation}

where

\begin{equation}
\begin{aligned}
N(r_c)= (2\pi^2 r_c^4 - 2\beta) \left(r_c^2-Q^2+3c \varepsilon r_c^{1-3\varepsilon}\right)-\alpha \pi r_c^2 \left(2Q^2-3c\varepsilon (3\varepsilon+1) r_c^{1-3\varepsilon}\right).
\end{aligned}
\end{equation}

Since the above equation is highly nonlinear, an exact analytical solution cannot generally be obtained. Therefore, we employ a perturbative approach by assuming that the thermal fluctuation parameters $(\alpha,\beta)$ and the quintessence normalization parameter $c$ are sufficiently small. Under this assumption, the corrected remnant radius can be expanded as

\begin{equation}
r_r=r_{r0}+\alpha r_1+\beta r_2,
\end{equation}

where $r_{r0}$ denotes the classical remnant radius, while $r_1$ and $r_2$ represent the first-order corrections induced by the logarithmic and higher-order thermal fluctuations, respectively.

The classical remnant radius is obtained by neglecting the thermal fluctuation terms $(\alpha=\beta=0)$, yielding

\begin{equation}
2\pi r_{r0}^{5}-2\pi Q^{2}r_{r0}^{3}+6\pi c\epsilon r_{r0}^{4-3\epsilon}=0.
\end{equation}

Discarding the trivial solution $r_{r0}=0$, the above equation reduces to

\begin{equation}
r_{r0}^{2}-Q^{2}+3c\epsilon r_{r0}^{\,1-3\epsilon}=0.
\end{equation}

For arbitrary values of the quintessence state parameter $\epsilon$, Eq.~(71) cannot be solved analytically. Nevertheless, considering the physically relevant regime $c\ll 1$, we write

\begin{equation}
r_{r0}=Q+\delta,
\end{equation}

where $\delta=\mathcal{O}(c)$ is regarded as a small perturbation.

Substituting Eq.~(72) into Eq.~(71) and retaining only the linear terms in $c$, one immediately obtains

\begin{equation}
\delta=-\frac{3c\epsilon}{2}Q^{-3\epsilon},
\end{equation}

which leads to the first-order analytical expression of the classical remnant radius,

\begin{equation}
r_{r0}=Q-\frac{3c\epsilon}{2}Q^{-3\epsilon}.
\end{equation}

Equation (74) shows that the presence of the quintessence field shifts the remnant radius from its Vaidya-Bonnor value by an amount proportional to the normalization parameter $c$. This classical solution constitutes the starting point for determining the quantum corrections induced by thermal fluctuations.

\subsection{Quantum-Corrected Remnant Radius}

Having determined the classical remnant radius, we now evaluate the effects of logarithmic and higher-order thermal fluctuations. Substituting the perturbative expansion

\begin{equation}
r_r=r_{r0}+\alpha r_1+\beta r_2,
\end{equation}

into the remnant condition $N(r_r)=0$ and retaining only the first-order contributions in the thermal fluctuation parameters, the corrected remnant equation can be decomposed into three independent parts corresponding to the classical, logarithmic and higher-order contributions. The classical part reproduces Eq.~(75), whereas the remaining terms determine the correction coefficients $r_1$ and $r_2$.

After straightforward algebraic manipulations, the logarithmic correction is found to be

\begin{equation}
r_1=
\frac{2Q^{2}r_{r0}^{-1}-3c\epsilon(3\epsilon+1)r_{r0}^{2-3\epsilon}}{10\pi r_{r0}^{4}-6\pi Q^{2}r_{r0}^{2}+6\pi c\epsilon(4-3\epsilon)r_{r0}^{3-3\epsilon}}.
\end{equation}

Similarly, the higher-order thermal fluctuation correction is obtained as

\begin{equation}
r_2=
\frac{\dfrac{2}{\pi}\left(r_{r0}-\dfrac{Q^{2}}{r_{r0}}+3c\epsilon r_{r0}^{-3\epsilon}\right)}{10\pi r_{r0}^{4}-6\pi Q^{2}r_{r0}^{2}+6\pi c\epsilon(4-3\epsilon)r_{r0}^{3-3\epsilon}}.
\end{equation}

Substituting Eqs.~(76) and (77) into Eq.~(75), the corrected remnant radius is obtained as

\begin{equation}
\begin{aligned}
r_r
=&
Q-\frac{3c\epsilon}{2}Q^{-3\epsilon}+\frac{\pi \alpha\left(2Q^3-3c\epsilon(3\epsilon+1)Q^{2-3\epsilon}\right)+{6\beta c\epsilon Q^{-3\epsilon}}}{2\pi^2 \left(2Q^4+3c\epsilon(4-3\epsilon)Q^{3-3\epsilon}\right)}.
\end{aligned}
\end{equation}

Equation (78) provides the analytical expression of the quantum-corrected remnant radius up to first order in the thermal fluctuation parameters. The first term corresponds to the classical Vaidya-Bonnor remnant modified by the surrounding quintessence field, whereas the remaining terms describe the effects of logarithmic and higher-order thermal fluctuations.

It is worth emphasizing that both correction parameters contribute differently to the remnant radius. The logarithmic correction, governed by the parameter $\alpha$, introduces a modification proportional to the black hole charge and the quintessence parameters. In contrast, the higher-order correction, controlled by $\beta$, becomes significant only through its coupling with the quintessence field. Consequently, the final remnant radius depends explicitly on the combined effects of the electric charge, the quintessence background, and the quantum thermal fluctuations.

In the classical limit $\alpha=\beta=0$, Eq.~(78) reduces to $r_r=r_{r0}$, thereby confirming the consistency of the perturbative solution.

\begin{figure}[htbp]
	\centering
	\includegraphics[width=0.75\textwidth]{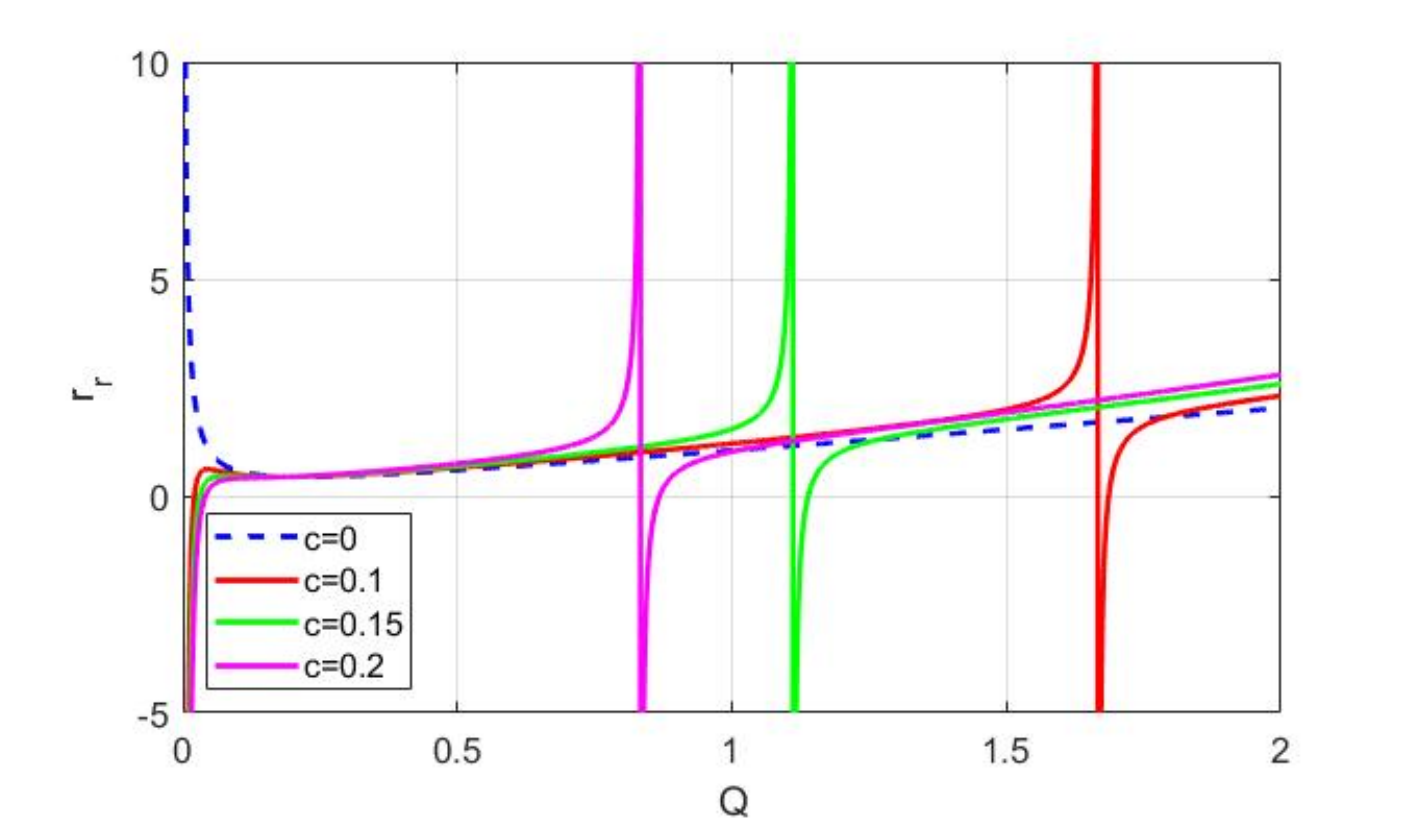}
	\caption{Effect of the  parameter $c$  on the remnant radius $r_r$.}
	\label{fig:r_c}
\end{figure}

Figure 23 shows the effect of the quintessence parameter $c$ on the remnant radius $r_r$. The remnant radius exhibits vertical divergences at critical charges. Increasing the quintessence parameter $c$ shifts the critical point toward lower values of $Q$ while enlarging the positive branch of $r_r$. This behavior demonstrates that quintessence promotes the formation of a larger and more stable remnant compared with the classical Vaidya-Bonnor black hole ($c=0$).

\begin{figure}[htbp]
	\centering
	\includegraphics[width=0.75\textwidth]{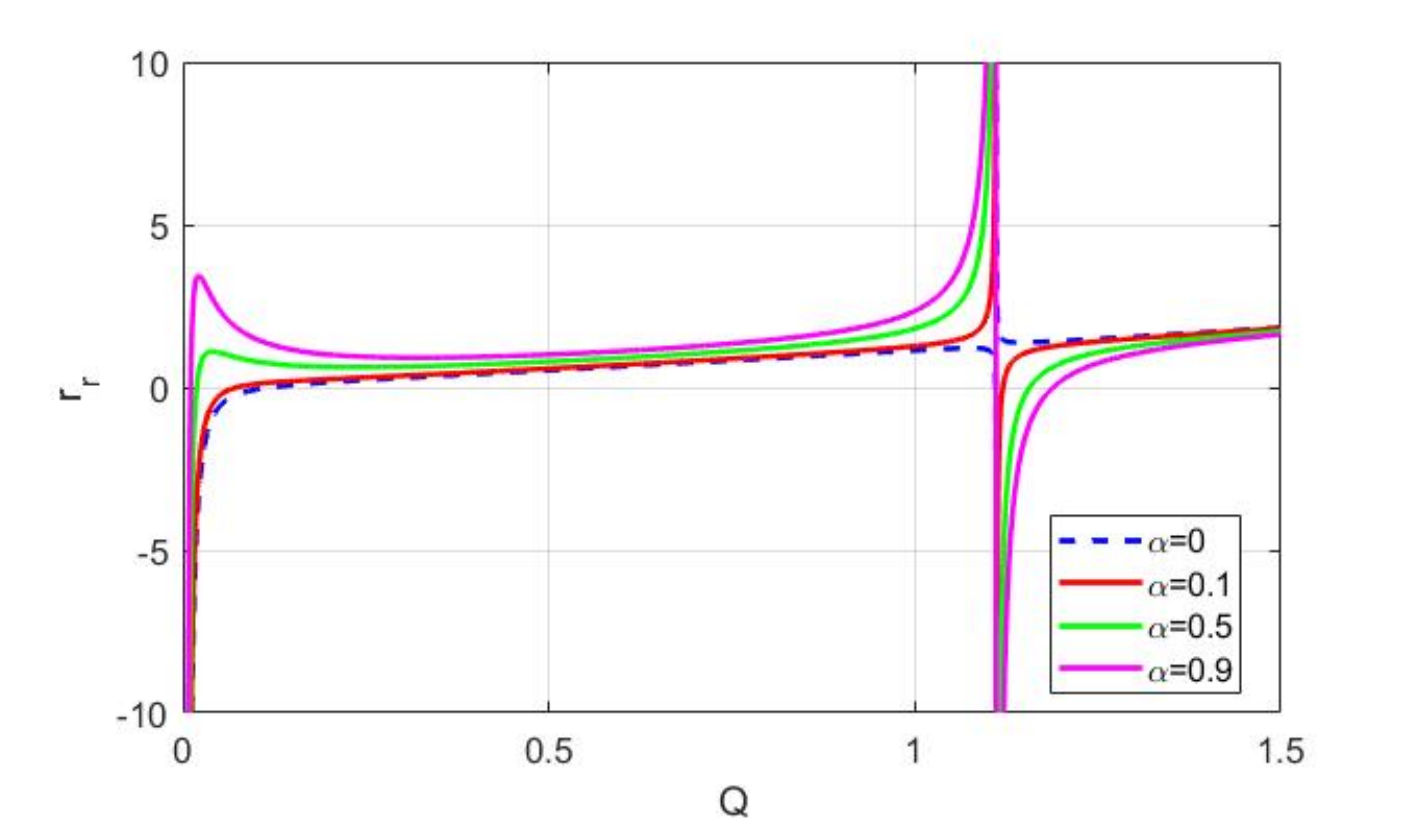}
	\caption{Effect of the logarithmic correction parameter $\alpha$ on the remnant radius $r_r$.}
	\label{fig:r_alp}
\end{figure}

Figure 24 illustrates the effect of the quantum correction parameter $\alpha$ on the remnant radius $r_r$. The remnant radius diverges at the critical charge. Larger values of $\alpha$ significantly increase the positive branch before the critical point and deepen the negative branch afterward, highlighting the strong influence of quantum effects. Compared with the classical solution ($\alpha=0$), first-order quantum corrections produce a larger, more stable black-hole remnant while shifting the critical behavior.

\begin{figure}[htbp]
	\centering
	\includegraphics[width=0.75\textwidth]{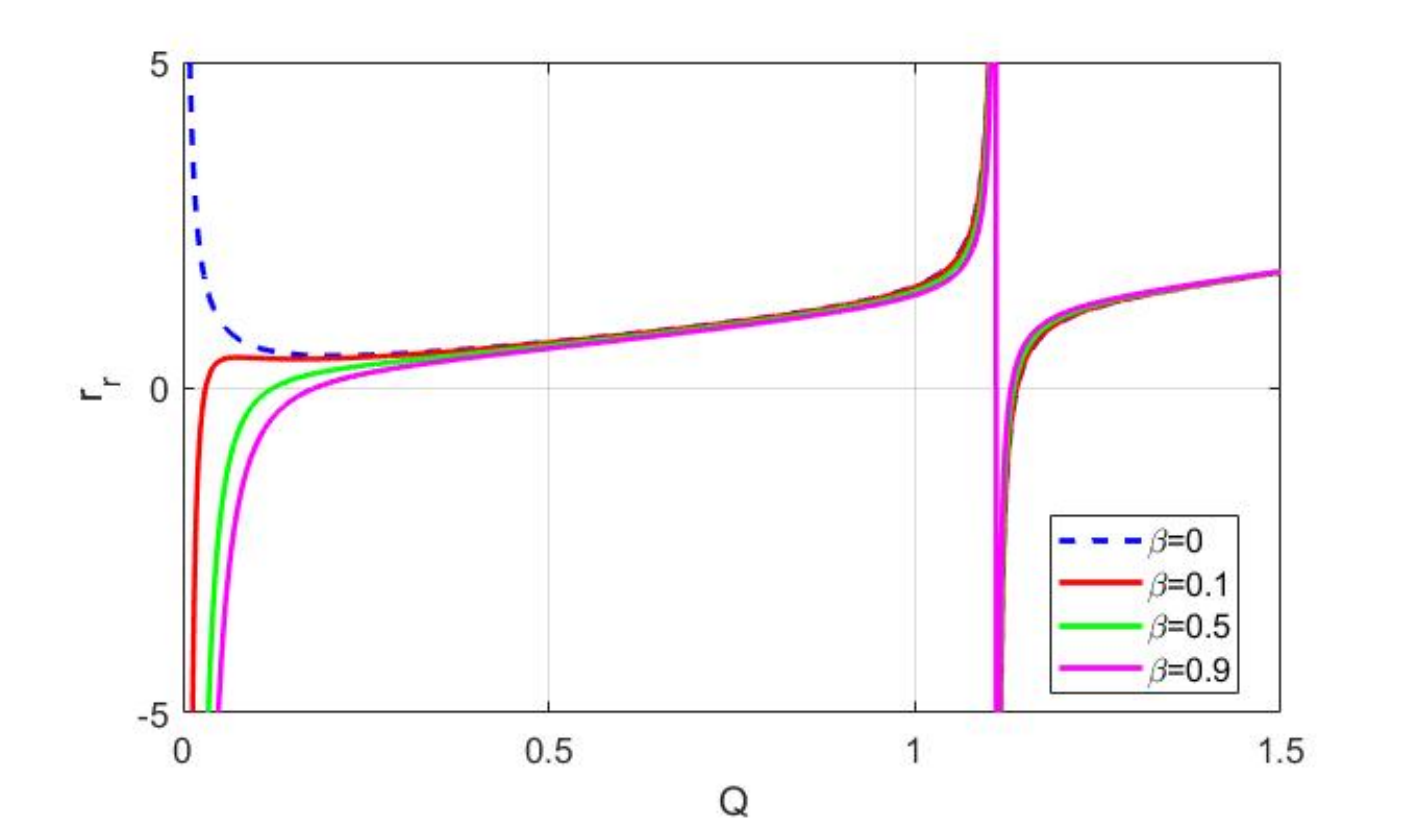}
	\caption{Effect of the quantum correction parameter $\beta$ on the remnant radius $r_r$.}
\label{fig:r_bet}
\end{figure}

As shown in Figure 25, the remnant radius undergoes a continuous divergence at a nearly fixed critical charge, confirming a second-order phase transition. Increasing $\beta$ extends the negative branch at small charges, whereas the positive branch remains almost unchanged away from the critical point. Relative to the classical case ($\beta=0$), higher-order quantum corrections mainly modify the remnant structure in the small-charge regime.

\subsection{Quantum-Corrected Remnant Mass}

Once the corrected remnant radius has been determined, the corresponding remnant mass can be obtained directly from the black hole mass function evaluated at the corrected horizon radius,[41-63]

\begin{equation}
M_r=\frac{1}{2}\left(r_r+\frac{Q^2}{r_r}-cr_r^{-3\epsilon}\right).
\end{equation}

Since the thermal fluctuation parameters are assumed to be small, the corrected remnant radius differs only slightly from the classical value. Consequently, the remnant mass can be evaluated perturbatively by expanding Eq.~(79) around the classical remnant radius and retaining only the first-order contributions.

Introducing

\begin{equation}
r_r=Q+\Delta,
\end{equation}

where

\begin{equation}
\Delta=-\frac{3c\epsilon}{2}Q^{-3\epsilon}+\alpha r_1+\beta r_2,
\end{equation}

the inverse radius and the quintessence term are expanded to first order as

\begin{equation}
\frac{1}{r_r}\simeq\frac{1}{Q}-\frac{\Delta}{Q^2},
\end{equation}

and

\begin{equation}
r_r^{-3\epsilon}\simeq Q^{-3\epsilon}\left(1-\frac{3\epsilon\Delta}{Q}\right).
\end{equation}

Substituting these expansions into Eq.~(79) and neglecting higher-order terms yields

\begin{equation}
M_r=Q-\frac{c}{2}Q^{-3\epsilon}+\frac{3c\epsilon}{2}Q^{-3\epsilon-1}\Delta.
\end{equation}

Finally, inserting the analytical expression of $\Delta$, the corrected remnant mass becomes

\begin{equation}
	\begin{aligned}
M_r=&
Q-\frac{c}{2}Q^{-3\epsilon}+\frac{3c\epsilon}{2}Q^{-3\epsilon-1}
\Bigg[-\frac{3c\epsilon}{2}Q^{-3\epsilon}+\frac{\pi \alpha\left(2Q^3-3c\epsilon(3\epsilon+1)Q^{2-3\epsilon}\right)+{6\beta c\epsilon Q^{-3\epsilon}}}{2\pi^2 \left(2Q^4+3c\epsilon(4-3\epsilon)Q^{3-3\epsilon}\right)}\Bigg].
	\end{aligned}
\end{equation}

Equation (85) represents the analytical expression of the quantum-corrected remnant mass. It is worth noting that the classical remnant mass is recovered in the limit $\alpha=\beta=0$, confirming the consistency of the perturbative expansion.

It is evident that the quantum corrections preserve a finite non-zero remnant mass even at the endpoint of Hawking evaporation. Therefore, the complete evaporation of the charged black hole is prevented by thermal fluctuations, leading to the existence of a stable remnant.

\begin{figure}[htbp]
	\centering
	\includegraphics[width=0.75\textwidth]{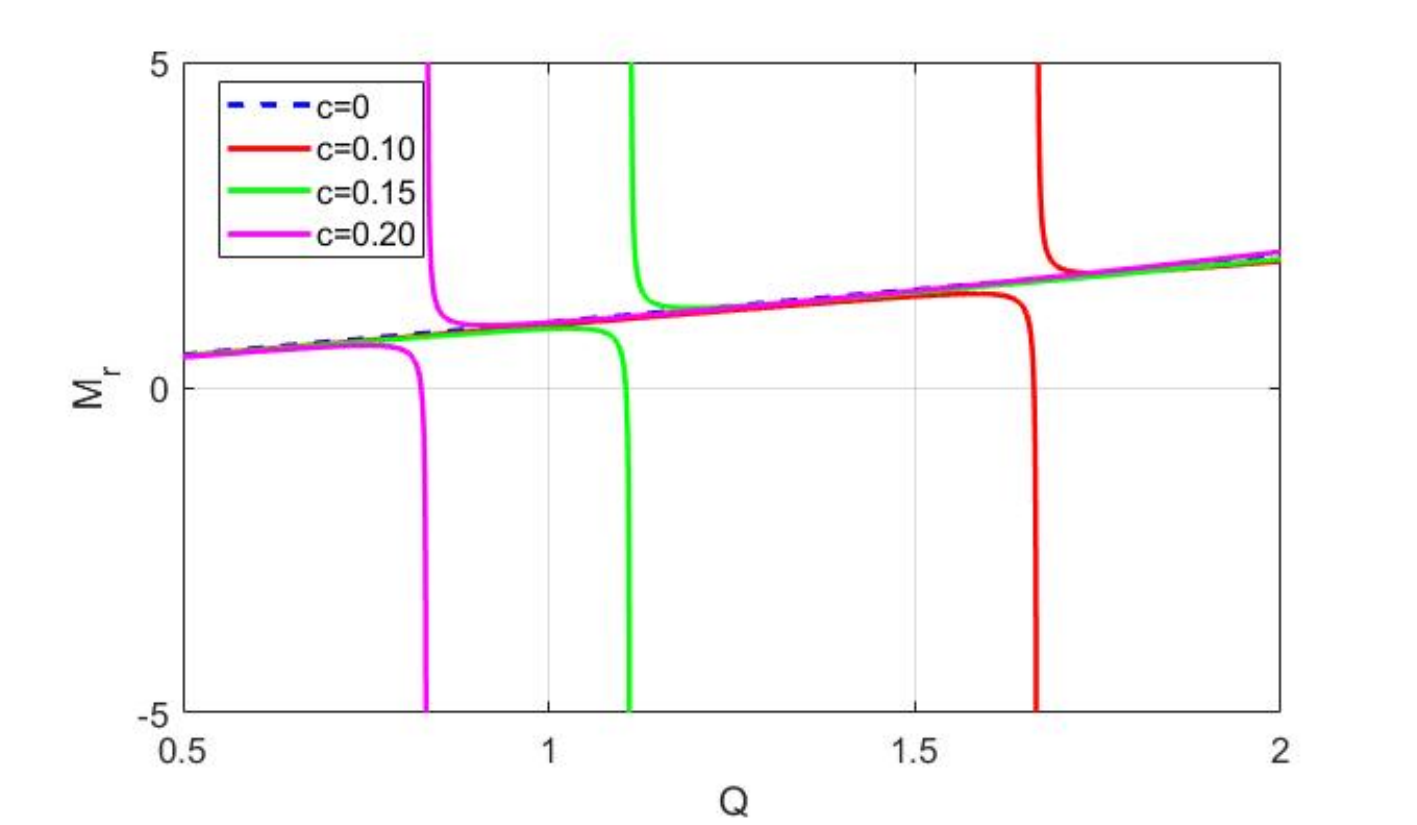}
	\caption{Effect of the  parameter $c$ on the remnant mass $M_r$.}
	\label{fig:M_c}
\end{figure}

As shown in Figure 26, the remnant mass $M_r$ increases smoothly with the electric charge except at critical points, where vertical divergences separate distinct physical branches. Increasing the quintessence parameter $c$ shifts the critical charge toward lower values while preserving a positive remnant mass away from the divergences. Compared with the classical case $(c=0)$, quintessence significantly modifies the remnant formation.

\begin{figure}[htbp]
	\centering
	\includegraphics[width=0.75\textwidth]{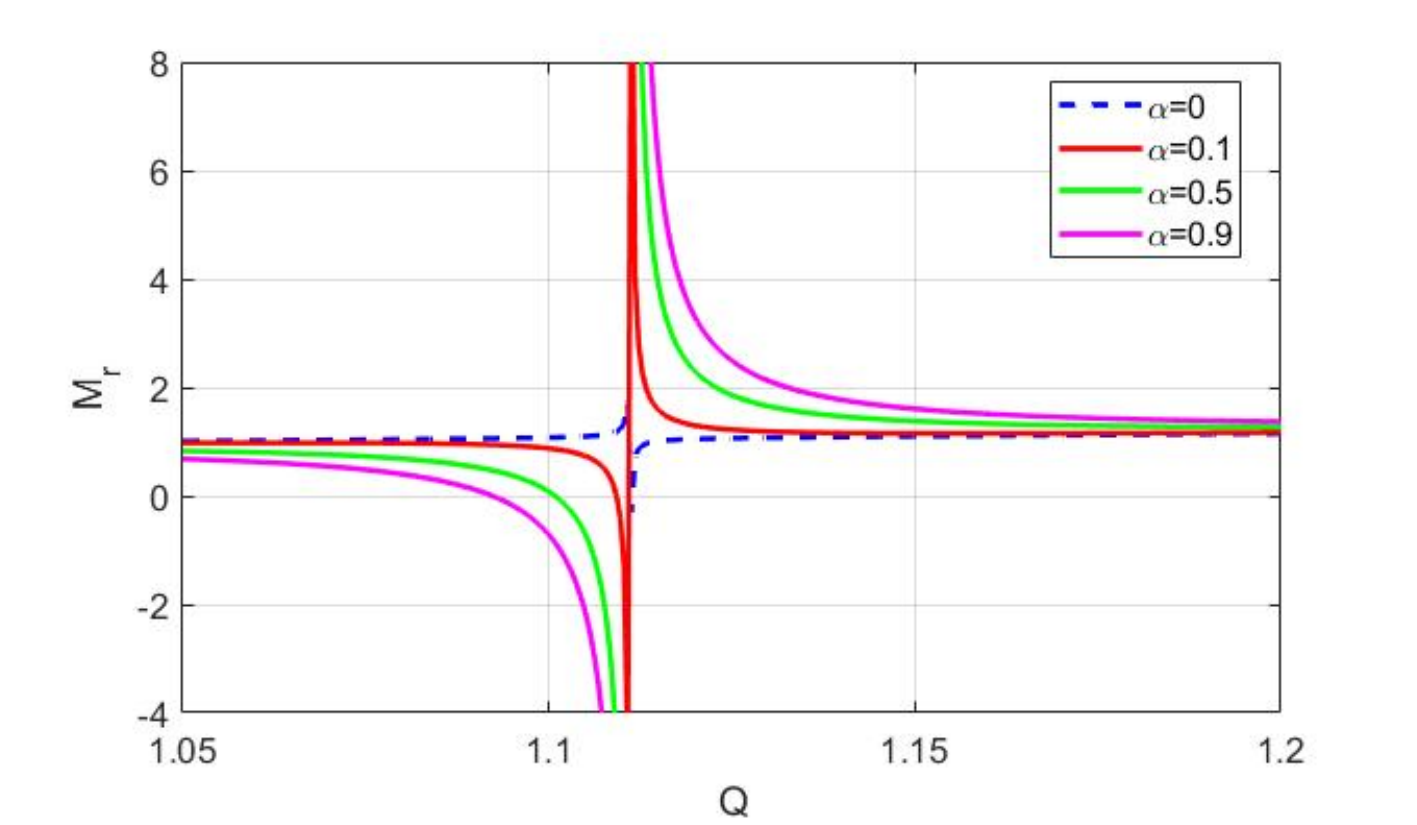}
	\caption{Effect of the logarithmic correction parameter $\alpha$ on the remnant mass $M_r$.}
	\label{fig:M_alp}
\end{figure}

In Figure 27, the logarithmic correction parameter $\alpha$ strongly affects the remnant mass near the critical charge. Increasing $\alpha$ enhances the positive remnant mass after the divergence and suppresses it before the critical point, while the critical charge remains almost unchanged. This indicates that logarithmic quantum corrections mainly modify the remnant mass without altering the critical configuration.

\begin{figure}[htbp]
	\centering
	\includegraphics[width=0.75\textwidth]{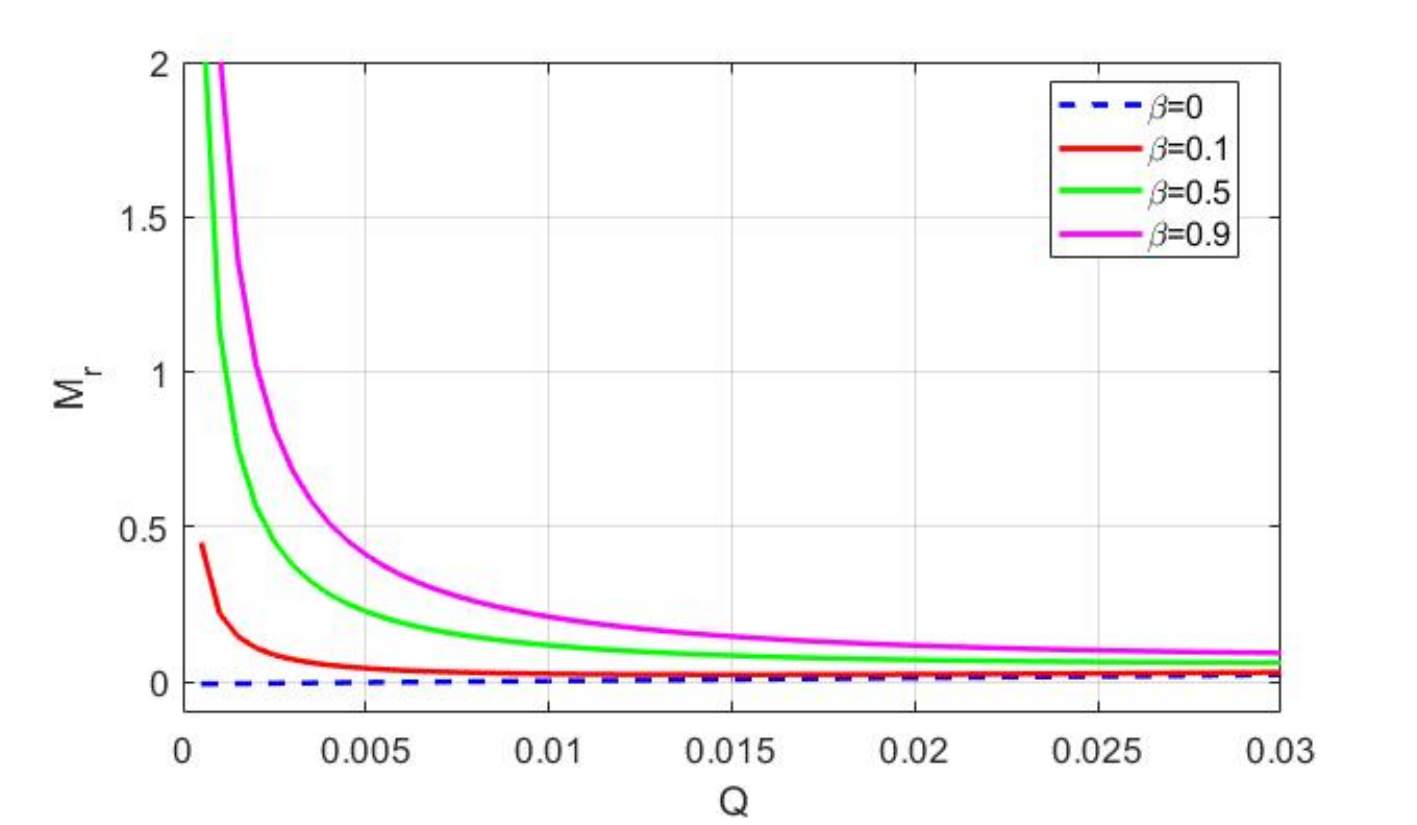}
	\caption{Effect of the quantum correction parameter $\beta$ on the remnant mass $M_r$.}
	\label{fig:M_bet}
\end{figure}

In Figure 28, the higher-order correction parameter $\beta$ increases the remnant mass over the entire charge range. Larger values of $\beta$ produce more massive remnants, especially for small electric charges, whereas all curves gradually converge at larger $Q$. Compared with the classical case $(\beta=0)$, higher-order quantum corrections enhance the stability of the final black-hole remnant.

\begin{figure}[htbp]
	\centering
	\includegraphics[width=0.75\textwidth]{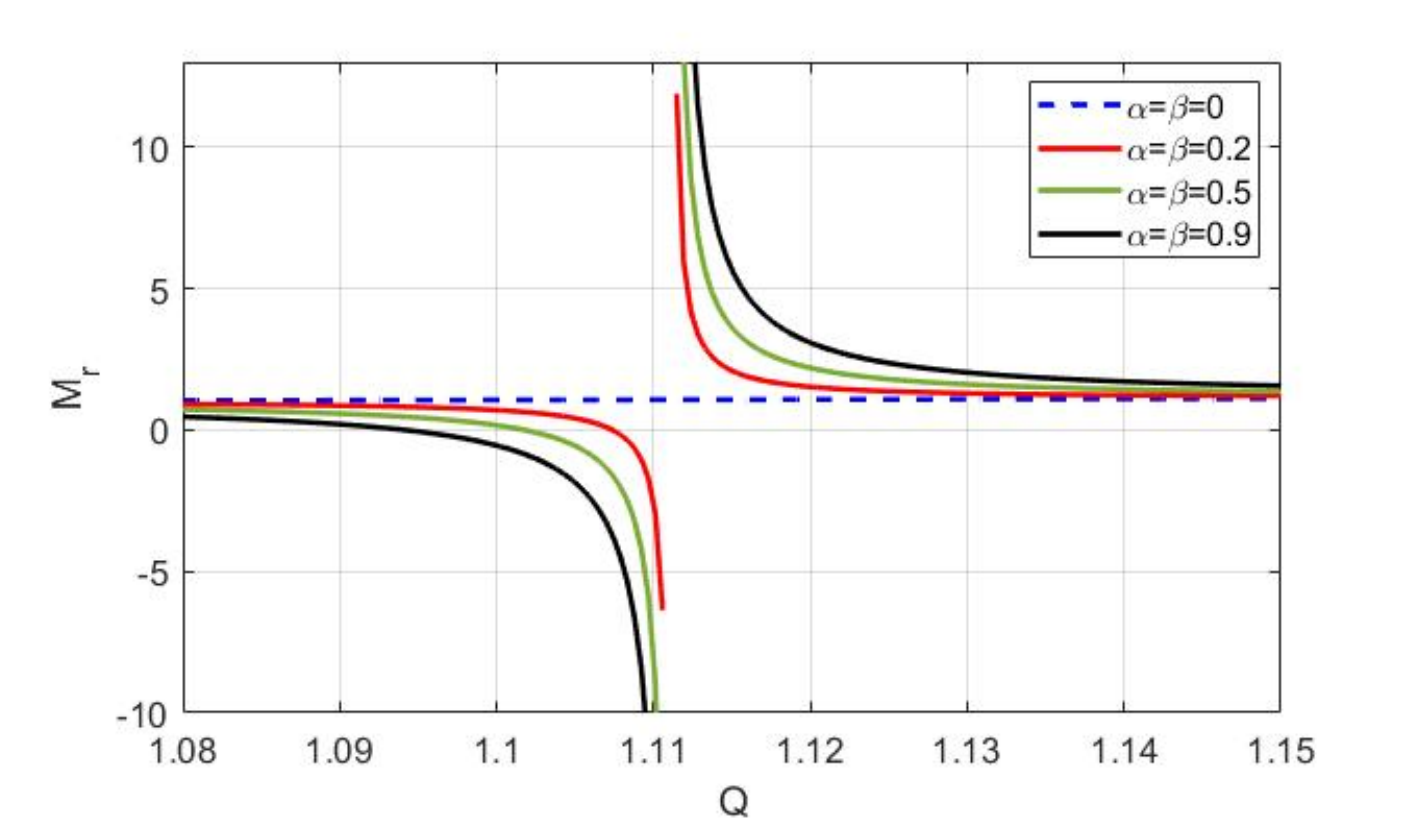}
	\caption{Effect of the logarithmic correction parameter $\alpha$ and the quantum correction parameter $\beta$ on the remnant mass $M_r$.}
	\label{fig:M_bet}
\end{figure}

As shown in Figure 29, the simultaneous increase of the quantum correction parameters $(\alpha=\beta)$ significantly modifies the remnant mass around the critical charge. The divergence position remains nearly unchanged, whereas the remnant mass increases on the stable branch and decreases on the unstable branch. These results demonstrate that combined quantum corrections enhance the persistence of the Vaidya-Bonnor black-hole remnant relative to the classical case.

The physically admissible remnant requires $r_r>0$, and $M_r>0$, while local thermdynamic stability additionally requires $C_c>0$.

These inequalities define the admissible region of the parameter space $(Q,c,\epsilon,\alpha,\beta)$ for which a stable quantum-corrected remnant exists.

The analytical expressions derived above clearly demonstrate that the remnant radius and remnant mass depend simultaneously on the electric charge, the quintessence parameters and the thermal fluctuation coefficients. Consequently, both the surrounding quintessence field and quantum thermal fluctuations play a crucial role in determining the final state of the black hole evaporation process.

\section{Conclusion}
In this work, we have investigated Hawking radiation and second-order quantum-corrected thermodynamics of a Vaidya-Bonnor black hole surrounded by quintessence. Using the Hamilton-Jacobi tunneling method, the radiation rate was obtained in Eq.~(35). The results in Figs.~1-4 show that the emission rate decreases with increasing horizon radius $r_c$, electric charge $Q$, and particle energy $E$, whereas increasing the quintessence normalization parameter $c$ enhances the tunneling probability. The equation-of-state parameter $\epsilon$ also significantly modifies the radiation through the radial dependence of the quintessence contribution. These results demonstrate that the electromagnetic field and the quintessence environment play opposite and complementary roles in regulating the Hawking emission.
The Hawking temperature, given by Eq.~(41), is reduced by increasing either $c$ or $Q$, as shown in Figs.~5 and 6. This suppression of the temperature indicates a slower evaporation process and shows that the surrounding quintessence and electromagnetic field substantially modify the thermal evolution of the black hole.
The second-order corrected entropy obtained in Eq.~(42) incorporates both the logarithmic correction parameter $\alpha$ and the higher-order correction parameter $\beta$. Figures~7-10 show that the quantum corrections become increasingly important as $r_c$ decreases. In particular, increasing $\alpha$ and $\beta$ enhances the corrected entropy, with the $\beta$ contribution becoming especially relevant near the final stage of evaporation. The corrected heat capacity and the other thermodynamic quantities derived from $S_c$ show that quantum fluctuations significantly modify the thermal stability and thermodynamic behaviour of the black hole in the small-horizon regime.
A central result of the present analysis is the formation of a finite remnant. The analytical expression for the corrected remnant radius, obtained in Eq.~(78), explicitly depends on $Q$, $c$, $\epsilon$, $\alpha$, and $\beta$, and reduces to the classical result when $\alpha=\beta=0$. Figures~23-25 show that quintessence and quantum corrections modify both the critical behaviour and the size of the remnant. Similarly, Figs.~26-29 demonstrate that the remnant mass is significantly affected by the quantum corrections, particularly near the critical charge. The higher-order correction parameter $\beta$ produces a pronounced increase of the remnant mass, especially for small $Q$.
For a physically admissible and locally stable remnant, the conditions $r_r>0,\qquad M_r>0,\qquad C_c>0$ must be simultaneously satisfied. Thus, the final state of the black hole is governed by the combined effects of the electric charge, quintessence field, and quantum thermal fluctuations. Overall, our results indicate that second-order quantum corrections provide an important mechanism for suppressing the final stage of evaporation and favouring the formation of a finite remnant. Further investigations could extend the present analysis by incorporating the backreaction of Hawking radiation on the Vaidya-Bonnor geometry, as well as by considering alternative quantum-gravity corrections. Such extensions would allow us to examine more deeply the robustness, stability, and physical properties of the remnant configuration obtained in the present framework. It would also be interesting to investigate how these corrections affect the evaporation rate, the thermodynamic behavior, and the energy distribution of the black hole surrounded by quintessence. Moreover, a study of linear perturbations and the associated quasi-normal modes could provide complementary information about the dynamical stability and relaxation properties of the spacetime.


\end{document}